**_Ab initio_ phenomenological simulation of the growth of large tumor cell populations**

Roberto Chignola[1,2], Alessio Del Fabbro[2,3], Chiara Dalla Pellegrina[1], Edoardo Milotti[2,4]

[1]*Dipartimento Scientifico e Tecnologico, Università di Verona, Strada le Grazie 15 – CV1, I-37134 Verona, Italy*

[2]*Istituto Nazionale di Fisica Nucleare, Sezione di Trieste, Via Valerio 2, I-34127 Trieste, Italy*

[3]*Facoltà di Scienze Motorie, Università di Verona, Via Casorati 43, I-37100 Verona, Italy*

[4]*Dipartimento di Fisica, Università di Trieste, Via Valerio 2, I-34127 Trieste, Italy*

**Running head:** Ab initio simulation of cell proliferation

**Correspondence to:** Roberto Chignola
Dipartimento Scientifico e Tecnologico
Università di Verona
Strada Le Grazie 15 – CV1
I-37134 Verona, Italy
Tel. +39 045 8027953
Fax +39 045 8027952
e.mail roberto.chignola@univr.it

**Abstract**

In a previous paper we have introduced a phenomenological model of cell metabolism and of the cell cycle to simulate the behavior of large tumor cell populations (Chignola R and Milotti E 2005 *Phys. Biol.* **2** 8-22). Here we describe a refined and extended version of the model that includes some of the complex interactions between cells and their surrounding environment. The present version takes into consideration several additional energy-consuming biochemical pathways such as protein and DNA synthesis, the tuning of extracellular pH and of the cell membrane potential. The control of the cell cycle - that was previously modeled by means of *ad hoc* thresholds - has been directly addressed here by considering checkpoints from proteins that act as targets for phosphorylation on multiple sites. As simulated cells grow, they can now modify the chemical composition of the surrounding environment which in turn acts as a feedback mechanism to tune cell metabolism and hence cell proliferation: in this way we obtain growth curves that match quite well those observed *in vitro* with human leukemia cell lines. The model is strongly constrained and returns results that can be directly compared with actual experiments, because it uses parameter values in narrow ranges estimated from experimental data, and in perspective we hope to utilize it to develop *in silico* studies of the growth of very large tumor cell populations ($10^6$ cells or more) and to support experimental research. In particular, the program is used here to make predictions on the behaviour of cells grown in a glucose-poor medium: these predictions are confirmed by experimental observation.

## 1. Introduction

Individual tumors are complex biological systems and, in spite of great therapeutic advances, many tumors still escape treatment and lead to death. Part of the complexity of the problem is a sheer consequence of tumor size: clinicians deal with the macroscopic properties of tumors, i.e. masses that may eventually weigh a few kilograms – and thus with a number of cells that ranges between $10^6$ and $10^{13}$ – and that may grow for months or years – with a corresponding number of cell cycles somewhere in the range between 100 and 10000. Microscopically, the malignant transformation of single cells is a multistep process that involves the modification of several molecular circuits which, in turn, modify the cells' behavior and the relationships between cells and the environment [1]. In addition, epigenetic and environmental factors – that include cell-cell interactions – also conspire with the bare genetic information to make tumor growth a highly variable process with very strong feedbacks [1].

The highly nonlinear character of the cells' internal molecular machinery, combined with the cell-cell and environmental interactions, with the large number of cells in a tumor, and with the extended tumor lifespan, make predictions based on the behavior of a single molecular circuit quite haphazard. We are thus in a frustrating situation, in which the huge body of detailed knowledge that has been collected in basic research areas, such as molecular biology, often seems to be of little or no consequence in the clinical management of the disease.

The availability of powerful computers has already helped bridge the gap between observations and predictions in many complex problems, and a few attempts have already been made to attack the problem of tumor growth with numerical models (see e.g. Refs. [2-4] and references cited therein). Recently we have proposed a numerical simulator of tumor growth [5, 6]: the simulator should eventually reproduce the growth of solid tumors in the prevascular phase and allow an *in silico* investigation of the biophysical laws that govern tumor growth dynamics and the response of cell clusters to anti-tumor treatments. The program simulates both tumor cells and the complex and changing environment where molecules, such as nutrients and drugs, diffuse and where cells interact mechanically with one another. The simulator deals with very different spatiotemporal scales and eventually with a large number of cells; using the prototype software we have been able to achieve a linearization of time and memory requirements using techniques and methods borrowed from molecular dynamics and computational geometry [5]. An essential part of the simulation software includes the code for metabolism and cell cycle, and in a previous paper we have shown and discussed the basic aspects of the underlying biophysical model [6]. Here we refine and extend that code and show that it can be used to simulate the proliferation of tumor cells that grow in a closed environment with good quantitative agreement with experimental data. We also use the model to predict the behavior of cells that grow in non-standard environments and compare these predictions with actual experimental data.

The next section briefly describes the materials and methods used in the experiments that have been carried out to fix some of the necessary parameters and in the numerical simulations. Section 3 contains a detailed description of the simplified metabolic network included in the simulator, and section 4 explains the implementation of the cell proliferation dynamics. The results of the simulations and comparisons with experimental data are given in section 5, while section 6 contains our conclusions and our outlook on future developments.

## 2. Materials and methods

### *2.1 Cell lines and proliferation assays*

Molt3 (human T cell leukemia) and Raji (human B cell leukemia) were obtained from the ATCC and maintained in RPMI-1640 medium supplemented with 10% foetal bovine serum. Cells were cultured at 37 °C in a 5% $CO_2$ humidified atmosphere and passaged weekly. For proliferation assays, $15x10^4$ cells from exponentially-growing cultures were seeded into T25 culture flasks in 7 ml of the following media: complete RPMI-1640 + 10% FBS or glucose-deprived RPMI-1640+10% FBS. Fifty 1 μl samples were drawn from culture flasks at the time point indicated in the figures to carry out measurements of cell survival. Alive and dead cells were identified using the vital dye Trypan-blue and were counted at the microscope using a Neubauer chamber. Four measurements for each well were carried out and the values were then averaged. Each data point in the growth curves is the mean ± standard deviation (SD) of three independent replicates. The amount of ATP in growing cell populations was measured using the luciferine/luciferase method (Promega kit) following the manufacturer's procedures. Measurements expressed in luminescence arbitrary units were converted to concentration units by means of a calibration curve obtained with known amounts of ATP. Luminescence was measured using a microplate luminometer reader (Bio-Tek instruments). Each data point in the figures is the mean ± SD of eight independent replicates.

Cell cycle analysis was carried out by flow cytometry. Cells were washed with PBS, and cell pellets produced from $1.5x10^5$ cells where resuspended under mild vortexing with 1 ml of cold Tris/HCl-EDTA buffer (0.1 M Tris-hydroxymethyl-aminoethane, 1 mM EDTA, pH 7.4). Cells were centrifuged at 4 °C and supernatants discarded. Pellets were labelled with 1 ml of a cold ethidium bromide solution (25 μM ethidium bromide, 0.3 mM Na-citrate, 1 mM NaCl, 0.1% v/v Triton X-100, 25 μg/ml RNAaseA) under mild vortexing, and cells incubated overnight at 4 °C . The nuclei were analyzed by flow cytometry on an EPICS-X (Coulter, Hialeah, FL) flow cytometer using a preparation from human peripheral blood mononuclear cells (blood drawn from healthy donors) as the standard.

*2.2 Simulations*

The simulation program has been written in ANSI/ISO-C++ and has been run on Apple computers (both PowerPC- and Intel-based). The program has been compiled with the GCC 4 compiler in a Mac OS X environment. The differential equations that make up the model have been integrated with a fixed time step $\Delta t = 1$ s; this choice is dictated by the conflicting needs of reducing the total computation time as much as possible and by the computational request - needed for algoritmic stability - that $\Delta t \ll \tau_k$ for all the Michaelis-Menten (MM) reactions, where $\tau_k$ is the reaction time constant for low concentrations for the $k$th MM reaction. The model is partly stochastic, and pseudorandom numbers have been generated by the linear congruent generator RAN2, taken from the Numerical Recipes library [7].

## 3. A minimal model of the biochemical network of tumor cells

The biochemical network implemented in the simulation program is shown in Fig.1: it is a really minimal network in comparison to the complexity of the actual metabolic system of tumor cells. The rationale behind the choice of the reactions in this minimal set has been discussed extensively in our previous paper [6], and we summarize again here the main arguments: 1. many parameters of the metabolic network are actually unknown; 2. even if we had a detailed knowledge of the actual metabolic network, a numerical model of a single cell's metabolic network would be computationally heavy, and the simulation of large cellular ensembles would be extremely impractical [5]; 3. many metabolic pathways are redundant and, in addition, it is now clear that the whole system has a hierarchical topology and that its dynamical behavior is dominated by the network's hubs [8]. We have thus decided to keep only those metabolic pathways that in our opinion do determine the cell's behavior: obviously this choice is somewhat arbitrary but it is eventually validated by the comparison with the experimental data.

The core of the metabolic network in Fig.1 contains the uptake and conversion of glucose and of other molecules into energy (in the form of ATP), storage molecules (e.g. glycogen) and waste products (e.g. lactate) and hence this part of the model is a global representation of glycolisis, oxidative phosphorylation through the TCA cycle and gluconeogenesis. Certain pathways are controlled by sensors, and we have considered both ATP and oxygen sensors. These sensors depend on the biophysical state of the virtual cell and tune its behaviour when ATP and/or oxygen become limiting factors because of harsh environmental conditions. All these aspects and their mathematical modeling have been described and discussed in our previous paper [6]. In the following subsections we concentrate on the new pathways that we have included to model growth in a closed environment. The values of model parameters for these pathways are listed in table 1. Since the environment is now closed it follows that we have to consider a conservation equation for the whole system's volume and a detailed description of the transport kinetics of some chemical species (like nutrient and waste products) inside and outside cells. But this, in turn, implies that the fate of the certain chemical species in cells must also be described and hence biosynthetic pathways that were previously neglected [6] must now be taken into consideration. We include also protein and DNA synthesis and since these paths consume energy the global energy balance of the cell must also be extensively revised as well.

*3.1 Closed environment*

The first significant biophysical difference with respect to our previous model is the definition of the closed environment where cells grow. In the previous version, in fact, cells were assumed to grow in an open volume where nutrients were always available and waste products immediately removed [6]. Now the environment is closed – as it happens to be in most experimental settings and in real tumors embedded in animal tissues – and environmental concentrations of nutrients, waste products, and other molecules become importantant variables. Because of the negligible compressibility of water solutions the environmental volume is effectively constrained by the equation :

$$Vol_{tot} = Vol_{env} + \sum_{k=1}^{N} Vol_k = \text{constant} \quad (1)$$

where $Vol_{env}$ is the volume of the extracellular environment and $Vol_k$ is the volume of the $k$-th cell.

*3.2 The pool of "other nutrients" in compartment A: glutamine as an essential component*

In our previous work [6], we introduced compartment A (see Fig.1) to include nutrients other than glucose that can nonetheless be catabolized through the oxidative phosphorylation pathways or converted into storage molecules through gluconeogenesis, and indeed it is known that tumor cell metabolism does not rely on glucose only. Since numerous but different molecules – each one involving different enzymatic mechanisms – can partecipate in cell metabolism, energy production and storage (in the form of glycogen, through the gluconeogenesis pathway that is described here phenomenologically by means of rate p11 – see Fig.1), we initially considered compartment A as a generic pool of "other nutrients" (e.g. lactate, glutamine and other aminoacids). However, in the present version of the model we also want to include biosynthetic pathways such as protein and DNA synthesis that are intimately linked to cell cycle kinetics (see the next paragraphs), and thus we need a better definition of the nutrients in the A compartment. We have decided to narrow the scope of the A compartment and focus on glutamine alone for the following reasons:

1. it is well known that glutamine is essential (together with glucose) to sustain tumor cell growth *in vitro* [9,10], and is required by real tumor cells *in vivo* [9,10], and it has also been shown that glutamine utilization for ATP production through the oxidative phosphorylation in tumor cells may actually overcome that of glucose [11];

2. it has been reported that most, if not all, tumor cells express the glycolytic isoenzyme pyruvate kinase type M2 (M2-PK) and that this enzyme is not expressed by normal cells [12]. The enzyme occurs in a highly active tetrameric form and in a dimeric form with low affinity for phosphoenolpyruvate. The switch between the two forms regulates glycolytic phosphometabolite pools and the interaction between glycolysis and glutaminolysis, the latter resulting in pyruvate and lactate production from glutamine [12]. Thus, these pathways do depend on glutamine (although the correlation depends on a series of enzymatic reactions) and this justifies the phenomenological description represented by rates p22 and p11, respectively (see Fig.1);

3. Finally, glutamine is an essential building block both of proteins and of DNA (see below) [13,14].

*3.3 Transport of nutrients and of waste products across the cell membrane*

Except for oxygen which diffuses freely across the cell membrane, transport of glucose, glutamine and lactate is mediated by saturable transporters whose activity is well known to follow a Michaelis-Menten type kinetics. Thus, if $X_{in}$ and $X_{out}$ represent the concentrations of a given molecule in the cell and in the environment, then transport across cell membrane is modelled by the equations:

$$\frac{dX_{in}}{dt} = -\frac{dX_{out}}{dt} = v_{in \to out} - v_{out \to in} \qquad (2a)$$

$$v_{out \to in} = \frac{v_{\max} X_{out}}{K_m + X_{out}} \qquad (2b)$$

$$v_{in \to out} = \frac{v_{\max} X_{in}}{K_m + X_{in}} \qquad (2c)$$

$$v_{\max} = v_m \cdot Surf \qquad (2d)$$

where $v_m$ and $K_m$ are the Michaelis-Menten constants for the molecular transport and *Surf* is the cell surface area. The transport of glucose depends on the oxygen concentration as well, and this dependence, along with the value of model parameters for glucose transport, has already been described [6].

*3.4 Intracellular and extracellular pH. Dependence of various transport and metabolic rates on environmental acidity.*

The intracellular and the extracellular pH are assumed to vary primarily because of lactate production. Lactic acid is produced by cells during the conversion of glucose through the anaerobic glycoslysis. During this conversion ATP is also produced, and from the stechiometry of ATP and lactic acid production we know that 2 moles of both species are produced per mole of glucose. Thus, the production rates of both species (see ATP_nOX in Fig.1) also match and have been parametrized by means of the rates $g_1$ and $r_1$ (see Fig.1) that describe the conversion of glucose-6-phosphate (G6P) and of glycogen storage molecules (STORE) through glycolysis (see [6] for further details).

Within the range of acidity that is physiologic for living cells, lactic acid is completely dissociated into $lactate^-$ and $H^+$ ions and these are co-transported through the cell membrane by means of MCT transporters. The MCT transporters do not require energy and transport proceeds bidirectionally along the lactate and proton gradients [15]. In a cell, protons partecipate to a complex series of chemical equilibria that regulate several important processes such as the cell osmotic pressure, the potential of the cell membrane and of the membranes of intracellular organelles (e.g. mitochondria), the transport of various ions across the cell membrane and the production of energy under the form of ATP. The regulation of the intracellular pH involves the action of energy-driven proton pumps ($H^+$ ATPase), proton channels and ion transporters that drive $H^+$ or equivalent $H^+$ and $HCO_3^-$ ions into and out of the cell and that are all expressed at the cell membrane [16]. The latter include an $Na^+/H^+$ antiport, $Na^+$-dependent $HCO3^-$ transporters, an $Na^+$-independent $Cl^-/HCO3^-$ exchanger, an $Cl^-/OH^-$ exchanger and other transporters such as the lactate-proton cotransporter [16]. In addition, a number of chemical reactions among which the set of reactions describing the buffering effect of the carbonate/bicarbonate system takes place both in the cellular cytosol and in the environment. Taken as a whole this system is rather complex (see e.g [17,18]) and includes fast chemical reactions, faster than the time interval of 1 s that we have taken as a compromise between algoritmic stability, computation time and computational load.

However some simplifications are possible: firstly, the experimental evidence shows that the buffering capacity of cytosol in tumor cells is much higher than that of the surrounding environment. Although the intracellular buffering capacity has been shown to vary slightly during the cell cycle, the intracellular pH does not vary much during the cell's life and this is due to the number of proton and ion pumps and of buffering systems that a tumor cell expresses and exploits for this purpose [19]. In addition, using a parametrization of the observed buffering capacity of cytosol [20], we carried out numerical simulations that demostrate that intracellular pH in simulated cells falls below physiologic limits only under severe starvation, that would anyway lead to cell death, e.g. because of insufficient ATP or glycogen storage levels. Thus, as a first approximation, we have assumed intracellular pH to be constant and we have fixed it at 7.2, which is the observed value in tumor cells [16,19,20]. Another simplifying consideration is that the buffering capacity of the environment can be measured quite easily, and for standard culture media the pH has been shown experimentally to be directly proportional to environmental lactic acid concentration [21]. In the simulator we take the following phenomenological model for the acidity of the extracellular environment:

$$pH_{out} = pH_{st} - \frac{AcL_{out}}{Vol_{out} \cdot \beta_{out}} \qquad (3a)$$

$$\beta_{out} = -\frac{dH_{out}}{dpH_{out}} \qquad (3b)$$

where $pH_{st}$ is the standard pH of a clean medium, $AcL_{out}$ is the mass of lactate in the extracellular environment, $Vol_{out}$ is the volume of the environment (and thus $AcL_{out}/Vol_{out}$ is the lactate concentration in the environment) and $\beta_{out}$ is the buffering capacity of the environment estimated from experiments [21].

The Michaelis-Menten parameter $v_{max}$ for the transport of glucose, lactate and of other molecules (see below) is known to depend on pH [22-25]. To the best of our knowledge, however, this dependence has not been studied in detail and no firm biophysical conclusions are available. Once again we resort to a phenomenological description that nonetheless takes into account the results of experimental observations. This description assumes that the $v_{max}$ of the various transporters is weighed by a function which is roughly a sigmoid curve as a function of pH. In practice we take the piecewise linear approximation for the weight function:

$$w(pH) = \begin{cases} w_1 & pH < pH_1 \\ w_1 + \dfrac{w_2 - w_1}{pH_2 - pH_1}(pH - pH_1) & pH_1 \le pH \le pH_2 \\ w_2 & pH > pH_2 \end{cases} \tag{4}$$

where $w_1, w_2, pH_1, pH_2$ are the parameters that we specify to approximate the few available data.

In particular, for the transport of glucose and of glutamine we take $w_1 = 0$, $w_2 = 1$, $pH_1 = 6.3$, $pH_2 = pH_{st} = 7.54$ (see [22,25]), while for lactate transport, we take $w_1 = 3$, $w_2 = 1$, $pH_1 = 6.0$, $pH_2 = 8.0$ [23,24].

It should be noted that since the intracellular pH is kept at a fixed value, then the flow of glucose, glutamine and lactate is unbalanced if the environmental pH differs from the standard value, and in the closed environment described by the simulator this does indeed happen because of lactate production and secretion by cells. As the environment is gradually more and more acidic, the uptake of nutrients is also reduced and can eventually switch off completely, thereby leading to a depletion of the energy reserves and ultimately to cell death. This mechanism, which is entirely mediated by the weight function described above, represents an important feedback regulatory circuit between cells and their environment, and can be tested experimentally because it defines the carrying capacity of the environment where cells grow (see the results section).

We take a fixed intracellular pH, however the increase of the concentration of $H^+$ ions in the cells cannot be neglected. The increase is originated by two main processes: as the cell population grows the lactic acid secretion also rises and the environmental pH lowers below the standard intracellular level: in this case a cell can import lactic acid through the MCT transporters. Secondly, the continuous intracellular production of lactic acid may eventually overcome the maximal rate of secretion of MCT transporters. We have already remarked that $H^+$ ions partecipate in a number of cellular processes in tumor cells, that collectively allow cells to maintain an almost fixed pH [16] and that these processes can be subdivided into two categories on the basis of their energy requirements. Since the energy consuming pathways are important in our model, we must take into account the consumption of ATP used to eliminate intracellular $H^+$ when the MCT transporters alone cannot sustain the whole load. It should be noted that we are not interested in the fate of the $H^+$ ions, because of the phenomenological modeling of the intracellular and extracellular pH, and we only need to focus our attention on the energy balance of $H^+$ transport.

To model the energy consumption due to $H^+$ transport, we recall that in the model $H^+$ ions originate from the complete dissociation of lactic acid, and that lactic acid is synthesized by cells through glycolysis and this path is described by rates $g_1$ and $r_1$ (see Fig.1; see also [6] for details). In addition lactic acid is transported out and in the cells by means of the MCT transporters with a transport kinetics that follows the set of equations (2a-d). $H^+$ can finally move across cell membranes by passive diffusion. The mass variation of $H^+$ ions in the cell therefore writes:

$$\frac{dH^+}{dt} = (2g_1 + 2r_1 + v_{out\to in,AL} - v_{in\to out,AL}) \cdot \frac{MW_{H^+}}{MW_{AL}} + D_{H^+} \cdot \left( \left[ H^+ \right]_{out} - \left[ H^+ \right]_{in} \right) \qquad (5)$$

where the factor 2 takes into accounts the stechiometric result that 2 moles of lactic acid are produced per mole of glucose, $v_{pAL}$ and $v_{mAL}$ are the rates of lactic acid transport in and out of the cells through MCT transporters, $MW_H{}^+$ and $MW_{AL}$ are the molecular weights of $H^+$ and lactic acid, respectively, $D_H^+$ is the effective diffusion constant (i.e. the diffusion coefficient multiplied by the thickness of the cell membrane) of $H^+$ in the cell membrane and $[H^+]_{out,\ in}$ are the extracellular and intracellular $H^+$ concentrations, respectively. Equation (4) is then used to calculate the energy consumption required to move the charged ions in the electric field associated to the Nernst membrane potential when this field opposes $H^+$ secretion (see below).

*3.5 Protein and DNA synthesis*

Protein synthesis and DNA synthesis are two important new blocks in the present simulation program, however the very high complexity of the actual cellular processes means that once again we have to resort to some form of approximate phenomenological description. In the case of protein synthesis we compute the number of glutamine and ATP molecules required to build an “average protein”. We take human serum albumin as a representative of the “average protein”; albumin contains 585 aminoacids and its molecular weight is ≈ 66.4 KDa [26]. Glutamine is incorporated in the protein as it is or after it has been changed into another aminoacid by enzymatic modification of its sidechain. Experimentally, it has been shown that in the proteins of HeLa tumor cells the aminoacids Ala, Asp, Glu, Gly, Pro, Ser derive from the glutamine carbon sidechain, although to a different extent [13]. For example, as much as 68% of Asp but only 3.2% of Ser derive directly from glutamine [13]. From these quantitative data and from the sequence of human albumin, we estimate that, on average, 5% of albumin aminocids are glutamine or are derived from glutamine. Thus, to synthesize one molecule of an average protein of 585 aminoacids the cell needs on average 29.25 glutamine molecules as building blocks.

An average protein of 585 aminacids contains 584 peptide bonds that require the hydrolysis of 2 molecules of ATP and 2 of GTP per bond for formation [27], and hence 1168 ATP molecules are utilized on the whole to synthesise one molecule of the average protein (note that in the present model the GTP molecular pathways are not yet taken into account).

Protein synthesis requires both glutamine and ATP as building blocks, and since either of them can vary in time and become a limiting factor, the rate of protein synthesis must depend on the building block which, at any given time, is less abundant (at the moment we neglect recovery of aminoacids due to protein degradation). From these considerations we find the equations that model protein synthesis:

$$\begin{cases} \dfrac{dPs}{dt} = \dfrac{p_{33}}{29.25} \cdot \dfrac{MW_{Ps}}{MW_A} & \textit{if} \quad \dfrac{N_{ATP}}{1168} > \dfrac{N_A}{29.52} \\ \dfrac{dPs}{dt} = \dfrac{v_p}{1168} \cdot \dfrac{MW_{Ps}}{MW_{ATP}} & \textit{elsewhere} \end{cases} \tag{6a}$$

$$p_{33} = p_3 \cdot d \tag{6b}$$

$$v_P = \frac{ATP_TOT \cdot ATPp}{K_{mP} + ATPp} \tag{6c}$$

where $P_S$ is the protein mass, $N_{ATP}$ and $N_A$ are the numbers of available ATP and glutamine molecules, respectively, $MW_{Ps}$, $MW_{ATP}$, $MW_A$ are the molecular weights of the "average protein", ATP and glutamine, respectively, $v_P$ is the rate of ATP pool consumption for protein synthesis (this is assumed to follow a Michaelis-Menten kinetics because of the activity of the enzymes that utilize ATP for protein synthesis), $ATP_TOT$ is the total ATP production (see [6] for details), $K_{mP}$ is the Michaelis-Menten constant for ATP consumption, $p_3$ is the rate of glutamine molecules channeled to protein synthesis and $d$ is a homeostatic control function that assumes real values in the interval [0,1] and that modulates the rate $p_{33}$ as a function of the size of the glutamine compartment A (see [6] for details). Initial guesses for $p_{33}$ and $v_p$ were obtained by considering that the protein content of one cell is approximately 10-20 % of the cell's mass [27], that a cell has a relative density ≈ 1 and that at the end of one cell cycle the protein mass must be approximately twice as much as in a newborn cell.

We have modeled DNA synthesis much like protein synthesis. However, in this case we note that the atoms $N_3$ and $N_9$ of purines and $N_3$ and $C_2$ of pyrimidines derive from the sidechain of glutamine after some biochemical processing [14, 27]. Thus, on average the stechiometric relationship between glutamine and the bases of the DNA is one molecule of glutamine for each base. We also note that the number of bases in the human genome is $6 \times 10^9$, and that the incorporation of nucleotides into the DNA sequence is sustained energetically by the hydrolysis of the high-energy phosphoryl groups present in the nucleotides themselves. The formation of the phosphodiester bonds (ΔG=5.3 Kcal/mol) – that join one base to the other – requires energy that is provided by ATP, then if we assume that during the duplication of the genome approximately $6 \times 10^9$ phosphodiester bonds are formed and that the hydrolysis of ATP releases ≈ 11.94 Kcal/mol [27], we can estimate that DNA duplication requires ≈ $2.67 \times 10^9$ ATP molecules. Thus both glutamine and ATP are required for DNA synthesis, and both may vary during the cell’s life and limit the DNA synthesis rate, which we model as follows:

$$\begin{cases} \dfrac{dDNA}{dt} = \dfrac{p_{33}}{6 \cdot 10^9} \cdot \dfrac{Nav}{MW_A} & if \quad \dfrac{N_{ATP}}{2.67 \cdot 10^9} > \dfrac{N_A}{6 \cdot 10^9} \\ \dfrac{dDNA}{dt} = \dfrac{v_{DNA}}{2.67 \cdot 10^9} \cdot \dfrac{Nav}{MW_{ATP}} & elsewhere \end{cases} \tag{7a}$$

$$v_{DNA} = \lambda_{DNA} \left(1 + \text{DNA_MAX_SPREAD} \cdot \xi\right) \frac{ATP_TOT \cdot ATPp}{K_{mDNA} + ATPp} \tag{7b}$$

where $v_{DNA}$ is the rate of *ATPp* consumption for DNA synthesis, *Nav* is the Avogadro constant, $\lambda_{DNA}$ is the fraction of the total ATP production rate *ATP_TOT* that is channeled to DNA synthesis in the S phase of the cell cycle, $K_{mDNA}$ is the Michaelis-Menten constant of the enzymatic process of ATP utilization for DNA synthesis, DNA_MAX_SPREAD is a constant, and $\xi$ is a random variable drawn from a uniform distribution on the interval (-1,1). The term DNA_MAX_SPREAD*$\xi$ is small cell-dependent random spread which is determined at cell birth and it has been introduced to parametrize the fluctuations assumed to occur in certain mechanical aspects of DNA synthesis that involve processes such as the unidirectional translocation of the motor proteins helicases to unwind and separate the two DNA strands [27]. Finally, it should be noted that we consider only one rate term for glutamine utilization for both protein and DNA synthesis. In this way the glutamine pool (compartment A in Fig.1) is subdivided into three pools that sustain the STORE compartment ($p_{11}$), energy production ($p_{22}$) and biosynthesis ($p_{33}$) and where $p_{11}$ and $p_{22}$ are dynamically interconnected through the ATP sensor (see also [6] for details).

*3.6 The energy balance: summary of the energy producing and consuming processes*

The model of cell metabolism takes into account diverse energy producing and consuming pathways. Energy, stored under the form of ATP, is produced by anaerobic glycolysis (rate *ATP_nOX* in Fig.1) and oxidative phosphorylation of glucose taken up from the environment (rate *ATP_OX* in Fig.1). In addition, if *ATP_OX* falls below a standard ATP production rate (e.g., under starvation conditions), a cell senses this difference and channels glycogen and glutamine catabolites toward oxidative phosphorylation to produce ATP (rates *ATP*2 and *ATP*3, respectively, in Fig.1). However, the production of glycogen molecules through gluconeogenesis has a cost that has previously been modeled by the rate *ConsATP* [6]. These paths have been discussed in detail earlier [6] and led us to define an overall production rate *ATP_TOT* and an overall balance between energy producing and consuming pathways, as in the following equations:

$$ATP_TOT = ATP_nOX + ATP_OX + ATP2 + ATP3 - ConsATP \tag{8a}$$

$$\frac{dATPp}{dt} = ATP_TOT - r_c \tag{8b}$$

where *ATPp* is the ATP pool and $r_c$ is a phenomenological ATP consumption rate. One of the main differences between the previous model and the present one is that now we define some of the energy consuming paths that were previously modelled by means of the single phenomenological rate $r_c$. In this context, we have already described above the use of ATP for protein and DNA synthesis. In addition we also consider a phenomenological rate of ATP consumption for mitochondria maintenance (i.e., maintenance of the mitochondria membrane potential which is essential for mitochondria to work properly [28], see $v_{Mit}$ in Fig.1) which is proportional to the number M of mitochondria in the cell. Finally, we include in the model the consumption of ATP used to pump $H^+$ ions outside the cell when the environmental pH becomes lower than the intracellular pH. The latter has been calculated as follows: we assume that, under normal environmental conditions, the cell has a membrane potential $V_{rif}$ = -21 mV (derived from calculations using the Nernst equation and assuming that the standard extracellular pH is 7.54 and the intracellular pH is 7.2, see below and [16]). When the concentration of environmental $H^+$ increases – because of lactic acid production and secretion – the membrane potential changes according to the Nernst equation:

$$V_{Nernst} = \frac{R \cdot T}{z \cdot F} \cdot \ln \frac{[H^+]_{out}}{[H^+]_{in}} \qquad (9)$$

where $R$ is the the gas constant, $T$ is the temperature (°K), $z$ is the proton charge and $F$ the Faraday constant. Under standard condition, the environmental pH is set to 7.54 and the intracellular pH to 7.2, so that $V_{Nerst}$=$V_{rif}$= -21 mV. As long as the environmental pH is greater than the intracellular pH, $H^+$ ions diffuse passively through the cell membrane. However, with a higher extracellulary acidity, $H^+$ ions cannot longer diffuse and must be actively pumped outside the cell: the energy required to pump $H^+$ ions outside the cells can be calculated from the Gibbs energy of ATP hydrolysis ($\Delta G_{ATP}$ = $3.1 \times 10^4$ J/mole) and the energy required to move a charged particle across the electric potential difference $\Delta V$=$V_{rif}$-$V_{Nernst}$. The rate of ATP consumption is:

$$v_{ATP_H} = H_Peff \cdot \frac{dH^+}{dt} \cdot \frac{F \cdot \Delta V}{\Delta G_{ATP}} \cdot MW_{ATP} \cdot \left(10^3 \frac{\text{mole}}{\text{kg}}\right) \tag{10}$$

where *H_Peff* is a constant that parametrizes the efficiency of the $H^+$ pumps, the rate $\frac{dH^+}{dt}$ has been defined in equation (4), $MW_{ATP}$ is the molecular weight of ATP and the multiplicative factor $10^3 \frac{\text{mole}}{\text{kg}}$ is the conversion factor that is needed to express the result in SI units (i.e., one mole of $H^+$ weighs 0.001 kg), so that $v_{ATP_H}$, the rate of ATP consumption for proton secretion against the electric potential, is expressed in kg/s as all the other rates.

Overall, the balance between energy production and consumption writes:

$$\frac{dATPp}{dt} = ATP_TOT - (v_P + v_{DNA} + v_{Mit} \cdot M + v_{ATP_H}) \tag{11}$$

## 4. Cell proliferation dynamics

In this section we describe how the model of cell metabolism is integrated with the biochemical network that drives the cell cycle and cell growth, and in particular we describe how we link the metabolic network to growth and division, i.e. the mechanism of molecular threshold implemented in the simulator. Next, we discuss the sources of randomness in the dynamics of cell division that are important in the simulator, and finally we list criteria for cell death.

The proliferation dynamics presented here differs in many ways from the dynamics in the previous version described in [6] and hence some preliminary comments are in order. First of all, the control of cell progression along the cell cycle was previously modeled by means of phenomenological thresholds. The cell cycle is conventionally represented by a series of phases characterized by specific molecular events and there are four major phases that cells traverse from birth until duplication: the initial growth factor-dependent phase G1, the S phase where DNA duplication takes place, the G2 phase during which a cell prepares its genetic material for proper sorting into the two daughter cells and the M phase were duplication (mitosis) occurs. Passage from one phase to the other is a unidirectional and highly coordinated process which is regulated in particular by a class of proteins called cyclins: the different members of the class are expressed at specific points along the cell cycle and activate kinases, the cyclin-dependent kinases (CDKs) [29]. Active CDKs, in turn, phosphorylate specific protein substrates often on multiple sites, and these phosphorylation events are tuned by a network of proteins, inhibitors and enzymes that phosphorylate or dephosphorylate the proteins involved, among which the CDKs themselves [29]. The result is a complex enzymatic network with feedback regulatory circuits that regulates the progression of a cell from one phase to the next at specific points called cell-cycle checkpoints [29]. This network has a fundamental relevance for both normal and tumor cell growth, and various models of its inner workings have been proposed [30].

The control of the progression along the cell cycle ultimately consists in the activation of kinases which phosphorylate substrates using ATP as the donor of high energy phosphoryl groups [29], therefore ATP depletion can be expected to block cell growth, and indeed the existence of two energetic thresholds at the G1/S and G2/M transitions has been demonstrated for tumor cells [31,32]. In our previous model we included only a rough description of these thresholds [6]. However, in a recent study of the dynamic properties of the multisite modification of proteins we have shown how this mechanism may apply to proteins that are central to the cell cycle checkpoint mechanism, and how it can set a biochemical threshold [33, 34]. The threshold obtained from multisite modification is robust and depends on the concentrations of all the molecules involved in the reaction, i.e., the checkpoints depend non only on the ATP concentration, but also on the concentrations of enzyme and substrate, and may better account for the observed variabilities of the cell cycle [33, 34].

A second major difference is the link between energy production and cell growth: the increase of cell volume was previously assumed to be proportional to the net uptake of glucose and to the overall rate of energy consumption [6]. The idea was that glucose uptake and storage provide a source of carbon atoms and that at the same time the cell mass increase can only proceed with a corresponding energy expense. Thus we considered a net balance of glucose uptake and storage given by a linear combination of the rates $v_{1p}$, $v_{1m}$, $g_1$ and $g_2$ [6], while energy consumption was modeled by means of the phenomenological rate $r_c$ discussed above. The rate $r_c$ described the global ATP consumption in processes such us protein and DNA synthesis, tuning and maintenance of the membrane potential and so forth, which, however, were not explicitly included in our previous model [6]. Unfortunately this description has a logical fault: ATP consumption depends on the amount of available ATP that in turn depends on cell metabolism where molecules such as glucose itself are converted into energy, and thus glucose uptake and ATP consumption, and hence energy consumption, are directly correlated, while they need not be. In addition, the hypothesis that cell volume increases as a function of energy consumption is biologically untenable since this does not consider possible dissipative pathways that are not related to cell growth, such as active secretion of newly synthesized proteins. Since we make explicit several energy consumption paths we must revise the link between cell metabolism and cell growth.

*4.1 Linking the metabolic network to the cell cycle*

In this subsection we summarize the conditions that each cell must satisfy to step from one phase to the next in the simulator, and we start with the most complex mechanism, which is responsible for the G1/S phase transition. The G1/S checkpoint is perhaps the best known and the most important one, because as soon as cells pass the checkpoint they become committed to progress along the remaining part of the cell cycle [35, 36]. The molecular details have been reviewed in several papers (see e.g. [35-40]).

Central to the underlying molecular network is the Rb/E2F complex. The retinoblastoma protein Rb (pRb) has 16 putative phosphorylation sites and may exist in various forms depending on the level of phosphorylation [37-40]. In its hyperphosphorylated form, when at least 10 sites are phosphorylated [37-40], the Rb/E2F complex is fully dissociated, but partial dissociation may occur for intermediate phosphorylation levels of the Rb protein. E2F is a transcription factor that once dissociated from the pRb starts the transcription of genes involved in a series of important pathways that ultimately bring about DNA synthesis [35-40]. This checkpoint, therefore, marks the end of the G1 phase and the beginning of the S phase. The phosphorylation of the Rb protein takes place thanks to the coordinated action of at least two different cyclin-dependent kinases (CDKs) that, in turn, are activated by cyclin D and cyclin E. These cyclins are expressed at appropriate phases of the cell cycle – in particular the expression of cyclin E is controlled by E2F after it is released by pRb upon partial phosphorylation – and form complexes with the specific CDK's. The CDKs' concentration in the cell is rather constant throughout the cell cycle, thus the timing of the CDKs' action is determined by the expression of the cyclins [29]. The enzymatic activity of the CDKs is also tuned by phosphatases and inhibitors that collectively form a complex biochemical network with feedback regulatory circuits [29, 30]. The network has already been modeled, and it has been shown that important dynamical behaviors emerge: thresholds, hypersensitive response and hysteresis [41]. In other words, the network reacts as an on-off irreversible switch [41].

Here we apply once again the approach that we followed in the case of cell metabolism: just as the metabolic network, the checkpoint control network has a hierarchical structure, and we assume its dynamics to be dominated by the system's hubs. Given its demostrated importance [35-40], we have selected the multiple phosphorylation process of the Rb protein as the central hub of the checkpoint control network. We have isolated this process from the network (Fig.2) and studied the general dynamical properties of the modifications of proteins on multiple sites (multisite protein modification, MPM)

[33, 34]. The results have been published recently and here we summarize the main conclusions:

1. MPM naturally produces a threshold in the system's response;

2. the threshold is robust to noise perturbation, and this is important when dealing with low protein concentrations, that is to say when fluctuations are not negligible;

3. if the protein that carries multiple modification sites controls a downstream Michaelis-Menten reaction (Fig.2), then MPM delays the downstream reaction and the delay may be several orders of magnitude larger than the characteristic times of enzyme kinetics. Thus MPM drives the information transfer from the fast kinetics of enzymes action to the slow kinetics of cellular response;

4. MPM dynamics does not depend on the attachment/detachment mechanism of the chemical groups (i.e. phosphoryl groups) that modify the protein with multiple modification sites and these may be represented by enzymatic processes or simplified as bimolecular interactions. It follows that MPM is equivalent to the classical allosteric effect [33, 34] and that, to model its dynamics, one can neglect the specific processes of protein modification.

The dynamical properties of MPM are rather attractive because they allow great simplification of the model of the G1/S checkpoint. In particular from property 3. we can represent the duration of the G1 phase in terms of an enzymatic reaction controlled by a protein with multiple modification sites and this is reminiscent of the way E2F drives the transcription of genes once released by the Rb control protein. Moreover properties 1. and 2. assure that the dynamical response behaves like a robust on-off switch, and property 4. justifies an approximate description.

The basic model for the control of the G1/S transition in our virtual cells is shown in Fig.2 and is based on the following assumptions:

1. the amount of ATP required to phosphorylate the Rb protein is negligible with respect to the overall amount of available ATP in the cell that is described in our model by the variable *ATPp*;

2. CycD and CycE instantaneously form complexes with specific CDKs whose concentration is constant during the cell cycle. As soon as the complex is formed the phosphorylation of the Rb protein starts;

3. the phosphorylation/dephosphorylation events of the individual available sites on the Rb protein follow a stochastic dynamics and are fast with respect to the observation time (i.e. with respect to CycD and CycE expression rates).

With these assumption we can describe the process of pRb phosphorylation by means of a probabilistic chain of phosphorylation/dephosphorylation events (transition chain) that occurs as a consequence of classical bimolecular reactions between cyclins and pRb [33, 34]. In particular, using assumption 3. we can neglect the dynamics of the transition chain and even the chain itself and concentrate instead on the equilibrium probabilities. The analysis in [33, 34] shows that if the E2F is released when at least $n_{th} = 10$ sites on pRb are phosphorylated, then we find the number of free E2F molecules:

$$P_{Rb} = N_{Rb} \cdot \sum_{n=n_{th}}^{N} \binom{N}{n} \cdot p^n \cdot (1-p)^{N-n} \tag{12a}$$

$$p = \frac{1}{2N[Rb]} \cdot \left\{ \left( N[Rb] + [Cyc] + K_{Rb} \right) - \sqrt{\left( N[Rb] - [Cyc] \right)^2 + 2K_{Rb}(N[Rb] + [Cyc]) + K_{Rb}^2} \right\} \tag{12b}$$

where $P_{Rb}$ is the number of activated pRb molecules with at least $n_{th}$ phosphorylated sites out of a total of $N$ sites, $N_{Rb}$ is the total number of pRb molecules, $K_{Rb}$ is the ratio between backward and forward rate constant of phosphorylation supposed to occur in a bimolecular reaction between the cyclin/CDKs complexes (CycD and CycE) and pRb, $[Cyc] = [CycD] + [CycE]$ is the total cyclin concentration, and where the square brackets denote concentrations.

A recent, more detailed analysis of the process dynamics has confirmed that it is safe to use the quasi steady-state assumption [34].

The number of activated pRb molecules depends on the concentrations of pRb, CycD and CycE. Using data in the current literature we assumed a phenomenological model for the kinetics of the three proteins as a function of the cell cycle phase [29, 35-40], and the results are shown in Fig.3. In this scheme, the Rb protein has been hypothesized to be synthesized in the G2 phase and then to be partitioned randomly at mitosis between the two daughter cells (see the next paragraph for a further discussion on this point). As far as we know, experimental evidence on the precise timing of the pRb expression is not available. Existing data indicate that the protein maintains an almost constant concentration in the cell but varies both its phosphorylation state and intracellular localization [29, 35-40]. Since cell division would result in a dilution of the pRb concentration the protein must be synthesized *de novo* sometime during the cell cycle of the mother cell. For the model described by the set of equations (12a-b) it is only important that pRb molecules are available for a new phosphorylation cycle in the daughter cells at the beginning of their life, hence the precise timing of pRb expression is irrelevant. On the other hand, CycD is synthesized at the beginning of the cell's life in the G1 phase and is destroyed immediately after the cell has overcome the G1m/G1p transition (see also Fig.2); CycE is synthesized in the G1p phase and is destroyed as soon as a cell enters the following S phase. The synthesis rate of these proteins is also assumed to be proportional to the overall protein synthesis rate defined by the set of equations (5). Thus, in general if *Xp* is the mass of either pRb, CycD or CycE:

$$\frac{dXp}{dt} = \alpha_{Xp} \cdot \frac{dPs}{dt} \tag{13}$$

The mechanism implemented in the simulator, whereby pRb releases E2F as soon as $n_{th}$ sites are phosphorylated and subsequently E2F catalyzes the conversion of a substrate S into a product R, models the whole set of enzymatic reactions that occur upon the release of the E2F in the dynamics of the G1m/G1p/S transitions. The pRb/E complex is assumed to follow a 1:1 stechiometry, thus the concentration of the enzyme E equals the concentration of active pRb molecules. This dynamics is described by the equations:

$$[E] = P_{Rb} \cdot [Rb] \tag{14a}$$

$$\frac{d[S]}{dt} = -\frac{k_{MM} \cdot [E] \cdot [S]}{K_{MM} + [S]} \tag{14b}$$

$$G1m \rightarrow G1p \quad if \quad [S] < S_{th,1} \tag{14c}$$

$$G1p \rightarrow S \quad if \quad [S] < S_{th,2} \tag{14d}$$

Where $[S]_0$ denotes the initial substrate concentration in the simulator, and $S_{th,1}$ and $S_{th,2}$ are two critical S concentrations. The initial substrate concentration $[S]_0$ is assumed to be the same for all cells. We remark here that the subdivision of the G1 phase into two subphases G1m and G1p has already been considered in our previous model [6] and is also a known experimental fact [42]. Here, however, this subdivision has a precise molecular meaning as it defines the transition to CycD destruction and CycE synthesis, and moreover both cyclin/CDKs complexes act on the same pRb substrate and this reflects realistically the present knowledge on the G1/S checkpoint [35-40].

At present, the molecular machinery of the G2/M checkpoint is more obscure than that of the G1/S checkpoint, and in that case we assume a much rougher threshold model, i.e., threshold crossing happens as soon as the concentration of the proper cyclin/CDK complex reaches a critical concentration. For this purpose, we consider a generic cyclin X that is synthesized at the beginning of the G2 phase and is destroyed after mitosis.

The M phase duration has small individual fluctuations associated to the mechanical aspects of chromosome condensation and sorting that take place during the M phase. Thus:

$$M_T = M_T_M\left(1 + \mathrm{PHASE_SPREAD}\cdot\xi\right) \tag{15}$$

where $M_T_M$ is the mean duration of the M phase, PHASE_SPREAD is a constant, and $\xi$ is a random variable drawn from a uniform distribution in the interval (-1,1). In the simulator, the fluctuation of the M phase is determined at cell birth.

The S phase is regulated differently: its duration is not determined by a checkpoint mechanism, but rather by DNA synthesis as described by the set of equations (6). The S phase ends when DNA completes the duplication process.

*4.2 Cell proliferation dynamics and stochastic aspects*

As a cell proceeds along the cell cycle its volume increases and all the cell material is roughly doubled in mass, so that after mitosis each daughter cell looks like the mother cell at birth. This also applies to mitochondria, and an approximately linear correlation between cell volume and the number of mitochondria has been demonstrated experimentally [43]. Mitochondria possess their own DNA that includes the genes for some, but not all, mitochondrial proteins. The other mitochondrial proteins are coded into the cell's DNA and imported from the cell cytosol after their expression. Mitochondria proliferate by fission, a process that is reminiscent of bacterial proliferation, and their number also varies because of mitochondrial fusion and death [44, 45]. The proliferation of mitochondria is highly coordinated with cell growth and the molecular signals at the basis of the cell/mitochondria synchronization are presently under investigation [44, 45]. The regulation of cell volume is not well understood at present, however in the simulator we decided to link the cell's volume to the number of mitochondria, i.e., we assume that each cell increases its size to adapt its volume to the number of mitochondria in the cytosol (and this part of the model has been completely revised with respect to the previous version [6]).

Mitochondria contain the molecular circuits where the oxydative phosphorylation of glucose takes place and produce ATP. On the other hand, the proliferation of mitochondria requires energy and hence it depends on the overall energy balance of the cell. This establishes an interesting interplay: a cell takes up nutrients from the environment for energy production which is the primary task of mitochondria; mitochondria utilize part of the energy for their maintenance and proliferation; finally, mitochondria give back energy to the cell to accomplish its various tasks.

These arguments lead to the following equations:

$$\frac{dM}{dt} = C1 \cdot \frac{dATPp}{dt} \tag{16a}$$

$$\frac{dVol}{dt} = C2 \cdot \frac{dM}{dt} \tag{16b}$$

where $M$ is the number of mitochondria, $Vol$ is the cell volume and $C1$ and $C2$ are two positive constants. It should be noted that unlike the previous version, the simulation program has no upper bound for cell volume and number of mitochondria, so that, at least in principle, they can now increase without limit. This means that volume and number of mitochondria are controlled by the metabolic network and by the growth dynamics, and the eventual stability hints at the correctness of the whole model.

At mitosis mitochondria are partitioned between the two daughter cells: the partitioning follows a binomial distribution, and the reasons for this have already been discussed [6]. The volume of the daughter cells is then calculated in accordance with equation (15b):

$$Vol = Vol_{\min} + C2 \cdot M_0 \tag{17}$$

where $Vol_{\min}$ is the volume of the nucleus [6] and $M_0$ is the number of mitochondria inherited from the mother cell.

The unequal sharing of cytosol in the daughter cells shows up as an additional source of randomness in the partitioning of pRb. The synthesis of pRb has been hypothesized here to occur in the nucleus during the G2 phase of the mother cell. In its hypophosphorylated form the protein is not released in the cytosol but is retained in the nucleus, bound to the nuclear matrix up to the early G1 phase [46], and thus the amount of pRb that the daugther cells receive depends on the distribution of the pRb molecules in the nucleus and on the dynamics of the nuclear division. In the simulator we implement this mechanism assuming that the nuclear matrix, and hence the associated pRb or any other such protein, is split according to a binomial distribution.

All the other molecules in the cytosol are partitioned proportionally to cell volume, and thus their concentration does not change at mitosis, and the randomness of their mass splitting depends on the fluctuation of the number of mitochondria.

*4.3 Criteria for cell death*

In the present version the model assumes the following criteria for cell death:

1. the ATP pool falls below a given threshold value $ATP_{\min}$. This may occur because of environmental nutrient deprivation;

2. the length of the S phase exceeds a given time span *S_T*. The length of the S phase in the simulator depends on the availability of ATP and glutamine, and thus on nutrient uptake and utilization. Experiments show that a long-lasting blockade of DNA synthesis is not compatible with cell life and the molecular mechanisms leading to cell death involve the biochemical networks that form the so-called intra S checkpoints [47-50];

3. the length of the G2 phase exceeds a given time span *G2_T*. This may occur in our model because of the G2/M checkpoint when [*CycX*] does not overcome the threshold $CycX_{th}$. Since [*CycX*] is a function of protein synthesis and cell volume, this process depends on nutrient availability and utilization as well. And indeed, experimental observations show that a long-lasting blockade of G2 phase progression leads to cell death [47-50].

## 5. Simulations vs. experimental results

### *5.1 Growth kinetics*

In its present configuration, the program simulates the growth of proliferating cells dispersed in a closed environment. This is equivalent to the growth of tumor blood cells in a tissue-culture plate and hence the results of simulations can be directly compared to observations. This comparison is all the more direct because the simulator uses parameter values estimated from actual experimental data.

On the whole, the model presented in [6] and extended here is quite complex because it considers many biochemical and cellular paths, and a true fit of the model parameters cannot be carried out, given the high number of parameters and the scarcity of some key experimental data. Thus the model is not optimized – although the parameter estimates all lie within reasonable biophysical extremes.

And yet, even in the absence of a true optimization, the model seems to be quite robust and is able to reproduce the observed patterns, e.g., Fig.4 and Fig.5 show some simulation outputs and experimental data on the growth of two human leukemia cell lines. The simulation covers the same timespan as the observations and there is good agreement between the simulated and the observed population dynamics, and the total computed ATP mass also fits quite well the experimental data. The only noticeable difference between simulations and experimental data is the fast decrease of the number of alive cells after the peak, and this probably means that the definition of cell death still lacks some relevant detail. However, cellular properties calculated for simulated cells using the parameters listed in Tab.1 and in Tab.1 in [6] - to which we refer as the "standard" parameters and that give the "standard" growth curve in Fig.1 - match fairly well those observed for real cells (Tab.2). Among them we find the cell size and the number of mitochondria per cell that, as mentioned earlier, have been modelled without built-in upper and lower bounds. On this basis, we conclude that the model is stable, and we discuss this topic further in the next section.

*5.2 Cells grow and modify the surrounding environment*

Growth curves such as those shown in Fig.4 and Fig.5 can be subdivided into two phases: an initial exponential growth phase followed by a decline in the cell number due to cell death. Both phases are well known to be related to the interplay between cells and their surrounding environment (see Ref. [51] and references cited therein). Initially, nutrients and space are abundant and we observe an exponential growth phase, however as cells grow they produce toxic waste products that accumulate in the environment ultimately leading to cell death in spite of nutrient availability. In ecological terms the size of the environment is related to the so-called carrying capacity [51, 52]. The interplay between cells and the environment has been studied both experimentally and theoretically by Tracqui et al. [52] using tumor cells that grow attached at the plastic surface of culture plates. The environmental accumulation of waste products, mainly lactic acid, has been monitored by these authors by means of pH measurements, and it has been shown that a peak in the growth curves corresponds to a rapid change in environmental pH towards acidic values which are no longer compatible with cell survival [52].

The data shown by Tracqui et al. can be simulated by the simulation program when we assume that cells die because of the decrease of environmental pH, and that this process does not depend on whether cells grow attached at the plastic surface of culture wells or in suspension. Fig.6 compares the result of a simulation with experimental data reported in [53] and redrawn here. To compare experiments with simulated data, time in Fig.6 has been rescaled in units of doubling time both for real and simulated cells. Doubling time is obtained by fitting the exponential growth phase with the equation:

$$N(t) = N(0) \cdot e^{kp \cdot t} \quad (18)$$

where $N(0)$ is the initial cell number and $kp$ is the growth rate. Then doubling time $\tau$ is calculated as follows:

$$\tau = \frac{\ln 2}{kp} \quad (19)$$

Fig.6 shows that our model is in good agreement with experimental data as far as cell behaviour and environmental changes are concerned.

*5.3 Simulations carried out under abnormal growth conditions: predictive capabilities of the model*

The model parameters listed in Tab.1 have been tuned with a lengthy procedure to simulate the growth of proliferating cells in an environment with standard nutrient concentration. Experimental growth media are supplemented with foetal bovine serum that provides hormones and various growth factors and contain a number of molecules that are required for cell growth such as vitamins, proteins, aminoacids, lipids and so forth. However, it is well known that tumor cells can grow even in poorer media, though they are sensitive to the deprivation of some key molecules that include glucose, glutamine and oxygen.

To further test our model we have simulated the growth of proliferating cells in an environment characterized by low glucose concentrations using the standard parameters listed in Tab.1. We developed the model without ever considereding such limiting growth conditions and hence these simulations are truly predictive. Fig.7 shows some results of these simulations. When the environmental glucose concentration is kept at the standard value of 0.9 $kg/m^3$, ATP production (per cell) remains roughly constant during growth. However, when the environmental glucose concentration is decreased, the model predicts a higher ATP production. Simulated cell in the model transiently produce more ATP in low glucose media because of the ATP sensor that has been fully described in our previous paper. The ATP sensor was introduced to tune the rates $r_3$, $p_{22}$ and $p_{11}$ shown in Fig.1 and was meant to model phenomenologically both the Pasteur and the Crabtree effects [6]. At low glucose concentrations, the ATP production rate falls below the standard assumed value (ATP_St, [6]) and turns on the ATP sensor. Additional STORE molecules and glutamine molecules are then catabolized resulting in a transient overshoot of ATP production (and oxygen consumption) [6].

As far as we know, this behaviour has never been explored experimentally and therefore we carried out experiments on tumor cell growth in glucose-poor media (see also the Materials and Methods section). RPMI glucose-free medium has been supplemented with 10% FBS that contains approximately 0.9 $kg/m^3$ glucose. Thus the glucose concentration in the whole medium is approximately 0.09 $kg/m^3$. Under these experimental conditions tumor cells can grow and the ATP production can be measured. Experimental data show that in standard medium the ATP per cell remains almost constant throughout the growth process whereas in low glucose medium the ATP production is significantly higher, in good agreement with the model predictions.

Next, we studied the effects of fast modulations of the environmental glucose concentrations on cell growth (Fig.8). Both simulations and experiments show populations that alternate periods of cell proliferation and death which are out-of-phase with glucose fluctuations, although the first peak in population size is higher for simulated cells. At the end of the first period of glucose deprivation, the model predicts an accumulation of cells in the S phase: this prediction is confirmed by the experimental observations.

The simulations shown in both Fig.7 and Fig.8 agree with the experiments, although the quantitative agreement is not perfect. We have carried out a rather extensive exploration of the parameter space and we believe that the origin is not due to some bad parameter but rather to the missing definition of some biochemical path that finely tunes the cells' behavior. Nevertheless, our model captures the major features of the cells' behavior in these unusual conditions.

## 6. Conclusions and outlook

When dealing with real complex systems, such as animal cells, with a large number of parts and a lot of redundancy, it is difficult to decide which parts of the system are really necessary and must be kept when developing a viable computational model. Several approaches can be followed - and indeed have been followed over the years - to model the growth of proliferating cells, and all have advantages and disadvantages. At one extreme are the analytical models that use coupled differential equations, and the major drawback of these approaches is that the discrete events, that mark a cell's life (such as the duplication of the genome and cell division), cannot be described because they lack the required analytical continuity. At the other extreme we find the models that try to approach numerically every single known molecular pathway, and still cannot describe all the cell's details because of the huge amount of computational resources required for this task [53, 54]. Our approach lies somewhere in between these extremes, and we have already shown that it can be used to model quantitatively some key aspects of cell metabolism and of cell cycle kinetics [6]. The approach is based on the fact that biochemical networks in the cell possess a hierarchical structure [8], and if a network has a hierarchical topology then the system dynamics is known to be dominated by the network's hubs [8]. Thus by modelling the hubs of the cell's biochemical networks one should – at least in principle – be able to capture most of the information of the cell dynamics. However, there are no general rules to identify these hubs and for a cell – with thousands of interconnected biochemical paths – the only way is to proceed heuristically.

A comparison between our previous model [6] and the model described here shows how model refinements lead to a better description especially when abnormal conditions require the turning on of seldom used molecular pathways. Indeed our previous model compared very favourably with many existing data in three related areas of cell biology: metabolism, growth and proliferation. However, that model was unable to account for deviations from standard growth conditions, because of the incomplete or missing description of certain important paths. The refined model presented here tries to overcome several previous limitations, and includes important new details:

1. the definition of a closed extracellular environment;
2. the description of biosynthetic pathways such as protein and DNA synthesis;
3. the description of a minimal biochemical network for the control of the cell cycle;
4. the description of additional energy-consuming paths;
5. the inclusion of stochastic aspects that are required to explain the observed fluctuations in tumor cell proliferation.

We believe that the introduction of a network that controls the cell cycle (point 3 above) is a major improvement in the model. Indeed, it is well-known that the control of cell cycle is rather complex as it is formed by a large number of proteins that act as enzymes and/or substrate with complex interactions that lead to a markedly nonlinear behaviour [29, 30, 35-41]. These aspects have been already been studied in detail both experimentally and theoretically and it has been shown that the network's dynamics includes emergent properties such as limit cycle oscillations and chaos [29, 30, 35-41]. However, these results have been obtained mostly in cell-free systems using protein cell extracts and it is not clear whether cells share the same dynamics. In particular, the chemical oscillations of some proteins of the cyclin network – although evident in the test tube – cannot be directly connected to cell cycle kinetics where proteins are shared by the daughter cells and where cyclins are destroyed and regenerated after gene expression: processes such as protein degradation and expression are sure to influence the dynamics of the network.

The mechanism that we have included in the simulation program, i.e., the multisite protein modification, and in particular multisite phosphorylation of the Rb protein, has been shown to account for at least three important, observed, dynamic behaviors:

1. generation of a biochemical threshold, required to allow a cell to proceed from one cell cycle phase to the other;
2. generation of a time delay in the downstream reactions that span several order of magnitude, that is required to transfer biochemical information from the fast enzymatic kinetics to the slow cell response;
3. stability of the dynamics even in presence environmental fluctuations.

The last remark means that Rb multisite phosphorylation opposes fluctuations, and tends to synchronize cells. Thus, the desynchronyzation of cell cycle kinetics, that is an experimental evidence [55] and that has been addressed in the previous model [6], is greatly reduced and must be restored with the introduction of some additional source of randomness. Our description of such processes is still at the phenomenological level, and yet we obtain realistic population growth curves (see Fig. 4 and Fig.5 as an example).

We have already remarked that the simulation program uses parameters that are estimated from actual data and that for this reason the results of simulations can be directly compared with observations. Here we add that during the simulation campaign that we have carried out to fine tune some parameter values, we have realized that most parameters are correlated and cannot assume arbitrary values. For example, if cells grow quickly because they express great amounts of cyclins or because they take up large amounts of nutrients, then they produce more waste products that lead to a fast increase of the toxicity of the environment. Thus the actual dimension of the parameter space is smaller than the total number of parameters, and parameter tuning is less complex than it appears to be at first sight.

Here we have shown that the model can reproduce quantitatively several aspect of the growth of large tumor populations of tumor blood cells and can be used to make testable predictions. We also wish to stress that the simulation program has already been able to simulate populations of more than a million cells: using a non-optimized version of the program we were able to simulate the growth of a cluster of cells that started from a single cell and exceeded 1.25 million cells after five days of processing using an Apple PowerMac G5. We believe that this marks an important milestone in the advancement of the simulation program and we hope that our model will eventually help the study of developmental features of tumor cell populations.

**Acknowledgement**

We wish to thank Professor Giancarlo Andrighetto for many useful discussions and for his enthusiastic support and encouragement.

## Figure captions

**Figure 1.** Schematic layout of the metabolic network that models cell metabolism and its relationship with the extracellular environment. Variables within circles represent molecular species and are expressed in units of concentration or mass. Symbols are as follows: Gex = external glucose concentration, Gin = intracellular glucose concentration, G6P = glucose-6-phosphate concentration, STORE = mass of glucose stored in the form of glycogen, Ac.Lat.ex = environmental lactic acid concentration, Ac.Lat.in = intracellular lactic acid concentration, Aex = environmental glutamine concentration, Ain = intracellular glutamine concentration, ATPp = pool of ATP molecules concentration, O2 = oxygen concentration, $H^+$ = mass of intracellular protons, Proteins = mass of intracellular proteins, DNA = relative mass of DNA (normalized to 1 for the whole genome). Rates are represented by squares. The dashed and dotted circuits represent the ATP and Oxygen sensors, respectively, and have been described previously [6]. The dashed-and-dotted circuit named "Cell Cycle Checkpoints" represent the molecular circuit of cell cycle control that has been modeled on the basis of previous studies on the dynamics of the allosteric effect (see Refs. [33, 34] and equations 11-14 for details). $\Delta\Psi$ is the cell membrane electric potential that has been used to calculate the energy costs for intracellular proton dissipation (see equations 8-9 for details). The calculation of the global energy cost also requires the rates v_Mit, v_P and v_DNA that model ATP consumption for mitochondria maintenance, protein synthesis and DNA synthesis, respectively.

**Figure 2.** Scheme of the molecular interactions that have been considered in the model of the G1/S transition of the cell cycle. The retinoblatoma protein (pRB) has been supposed to form a complex with an enzyme E and carries 16 putative phosphorylation sites. The pRB-E complex is synthesized in the G2 phase and it is partitioned at random between the daughter cells at mitosis. At the beginning of the G1m phase the cyclin D protein (CycD) is expressed and it phosphorylates the pRB-E complex upon rapid association with specific cyclin-dependent-kinases [35-40]. Phosphoryl groups added to the pRB-E complex are represented graphically by means of black circles. The phosphorylation event is assumed to occur following the reversible bimolecular interaction between cyclin D and the pRb complex: a detailed study has shown that the precise mechanism of pRb phosphorylation is irrelevant with respect to the dynamics of the system's response [34]. Upon partial phosphorylation of the pRb, a fraction of E molecules are released and catalyze a reaction whereby a substrate S is converted into a product P. When the concentration of S falls below the threshold $S_{th,1}$ (see equations 13) two paths are activated the expression of the cyclin E protein (CycE) is activated and this marks the progression of the cell from the G1m phase to the following G1p. CycE participates to pRb phosphorylation upon istantaneous association with specific cyclin-dependent-kinases and leads to the hyperphosphorylation of the pRb with full E detachment. When the concentration of S falls below the threshold $S_{th,2}$ (see equations 13) the cell progresses to the S phase. At the end of the G1m and of the G1p phases CycD and CycE, respectively, are destroyed by proteolytic degradation that is assumed to occur faster than the considered reaction kinetics. The enzymatic reaction catalized by E is assumed to model phenomenologically the reactions controlled by the transcription factor E2F that associates to pRb in real cells. The nature of pRb

phosphorylation naturally introduces a threshold and a time delay in the reaction catalized by E with characteristic times comparable with the cell progression through the various cell cycle phases (see also [33, 34] for a detailed biophysical description).

**Figure 3.** Simulation of cell cycle kinetics and its relationships with the expression of key molecular components of cell cycle control. A: kinetics of the cell cycle transitions from one phase to the other for a simulated cell. Cell cycle phases are as follows: 1 = G1m, 2= G1p, 3 = S, 4 = G2 and 5 = M. B-E: these graphs show the variation of cyclin X, D and E and of the pRb during the same time span as panel A (see text). Finally, panel F shows the fraction of hypersphosphrylated pRb that has been calculated using equations (11). Notice that in the present model the oscillations in cyclin expression (and of other molecular components of the cyclin network), that have indeed been observed experimentally, are the result of protein expression and degradation in successive cell cycles rather than the result of complex nonlinear interactions between molecules. As far as we know, no models of cyclin oscillations have taken into account the intermittent protein expression and degradation through successive cell cycles as the driving force of their oscillatory behaviour. More realistically, the two aspects are probabily integrated in real cells.

**Figure 4.** Simulation of cell growth and death in a limited environment. In both panels, data are expressed as cell density vs. time. Upper panel: simulation outputs for different values of the parameters described in this paper. The curves marked as run120 show the simulation outputs obtained with the parameter values listed in Tab.1 and in Tab.1 of Ref. [6] and which we call “standard”. Lower panel: experimental growth curves

measured for the two human leukemia cell lines MOLT3 and Raji following the procedures described in the Experimental section.

**Figure 5.** Amount of the ATP pool in simulated and real cells in growth assays. Upper panel: simulation outputs. Symbols are the same as in Fig.4 for proper comparison. The grey dashes represent the SD calculated around mean values for the simulated populations. Lower panel: experimental results obtained with the human leukemia cell lines MOLT3 and Raji following the procedures described in the Experimental section.

**Figure 6**. While cells grow they modify the surrounding environment. This picture shows the results of simulations compared to actual data described in [52], and the relationships between cell concentration and environmental pH at different times during the growth assay. In both panels, the cell density has been normalized with respect to the initial cell concentration C(0). Time has also been rescaled in units of cells' doubling time to compare the data obtained with fast growing simulated cells and slow growing real tumor cells (see the text for details). Upper panel: simulation outputs. Lower panel: experimental results redrawn from [52].

**Figure 7.** Predictive potential of the model: ATP production by cells growing in low glucose environments. A: amount of the ATPp calculated for cells in simulations carried out with the parameters' values listed in Tab.1 for different concentrations of external glucose. The symbol +G refers to the standard glucose concentration of 0.9 $kg/m^3$, which is the typical glucose concentration in complete culture media supplemented with FBS. After the startup phase of the program, that allows the concentrations of all the considered molecular species to settle at equilibrium values, the environmental glucose concentration has been reduced to 10% or to 7.5% of the standard value. B: experimental resulta obtained with MOLT3 cells in complete medium (+G) and in two independent experiments where cells have been grown in glucose-free RPMI medium supplemented with 10% foetal bovine serum (FBS). Glucose is present in FBS with an approximate concentration of 0.9 kg/m$^3$, but this may vary in different FBS batches. Thus in these two experiments the final environmental glucose concentration is approximately 0.09 kg/m$^3$. C and D: cumulative ATP for simulated cells (panel C) and real cells (panel D) (the cumulative value has been calculated from the curves in panels A and D with a simple integration: $\int_0^\infty ATP(t) \cdot dt$).

**Figure 8.** Predictive potential of the model: growth of tumor cells in environments subjected to cycles of glucose deprivation. A and B: Time-dependent variations of glucose concentrations in the environment of simulated cells (A) and of real cells (B). The concentration of environmental glucose for real cells has been estimated as described in the caption of Fig.7. C and D: growth curves (cell density vs. time) for simulated (panel C) and real (panel D) cells. E: cell cycle phase distribution for simulated cells during the first cycle of environmental glucose deprivation. Simulation outputs show an accumulation of cells in the S phase and a strong reduction of cells in the G2/M phases. F: raw flow cytometry data of cell cycle distribution of real cells during the first cycle of environmental glucose deprivation. Data have been collected as described in the Experimental section. The accumulation of the cells in the early S phase is evident (arrow) as well as the reduction of the cells in the G2/M phases,

although both data sets do not match quantitatively the simulation outputs. However, the qualitative pattern is well predicted by the model.

**Table 1.** Model parameters for the *standard* simulation that are either modified or new with respect to those listed in reference [6].

| Symbol | Value | Units | Meaning | References |
|---|---|---|---|---|
| $V_{max_A}$ | $1\times10^{-9}$ | $kg\ s^{-1}\ m^{-2}$ | Maximum rate of glutamine transport | [12] |
| $K_{mA}$ | 0.0238 | $kg\ m^{-3}$ | Michaelis-Menten constant of glutamine transport | [56] |
| $V_{max_AL}$ | $9.58\times10^{-8}$ | $kg\ s^{-1}\ m^{-2}$ | Maximum rate of lactic acid transport | [15, 57, 58] |
| $K_{mAL}$ | 0.4053 | $kg\ m^{-3}$ | Michaelis-Menten constant of lactic acid transport | [15, 57, 58] |
| $V_{rif}$ | -0.021 | V | Cell membrane potential in standard conditions | This paper (see the text for details) |
| *H_Peff* | 0.1 | | Phenomenological constant describing the inverse of the efficiency of proton pumps | This paper (by data fitting) |
| $D_H$ | $1.5\times10^{-18}$ | $m^3\ s^{-1}$ | Effective proton diffusion constant (assuming a plasma membrane thickness of 10 nm) | [59] |
| $\beta_{out}$ | 0.19953 | $kg\ m^{-3}$ | Buffering capacity of the environment | [21] |
| $p_3$ | $7\times10^{-20}$ | $kg\ s^{-1}$ | Rate of glutamine consumption for protein and DNA synthesis | This paper (by data fitting) |
| $K_{mP}$ | $9\times10^{-17}$ | kg | Michaelis-Menten constant of ATP utilization in protein synthesis | This paper (by data fitting) |
| $K_{mDNA}$ | $8\times10^{-18}$ | kg | Michaelis-Menten constant of ATP utilization in DNA synthesis | This paper (by data fitting) |
| $\lambda_{DNA}$ | 0.008 | | Fraction of the total ATP production rate utilized for DNA synthesis | This paper (by data fitting) |
| *DNA_MAX _SPREAD* | 0.1 | | Maximum spread of the fluctuations in the DNA synthesis duration | This paper (by data fitting) |
| N | 16 | | Maximum number of phosphorylation sites on pRb | [38] |
| *l* | 10 | | Number of sites that must be phosphorylated for pRb activation | [38] |
| $K_{Rb}$ | $10^{-6}$ | $M^{-1}$ | Equilibrium constant for | [60] [a] |
| | | | phosphorylation/dephosphorylation process on pRb | |
| $\alpha_{pRb}$ | $1.5x10^{-2}$ | | Fraction of proteins which is pRb | This paper (by data fitting) |
| $\alpha_{CycD}$ | $7x10^{-3}$ | | Fraction of proteins which is cyclin D | This paper (by data fitting) |
| $\alpha_{CycE}$ | $10^{-3}$ | | Fraction of proteins which is cyclin E | This paper (by data fitting) |
| $\alpha_{CycX}$ | $8x10^{-3}$ | | Fraction of proteins which is cyclin X | This paper (by data fitting) |
| $S$ | $10^{-3}$ | M | Substrate concentration for the downstream Michaelis-Menten reaction controlled by pRb | [33, 34] |
| $k_{MM}$ | $10^{4}$ | $s^{-1}$ | Rate of *S* molecules consumption | [33, 34] |
| $K_{MM}$ | $10^{-3}$ | M | Michaelis-Menten constant of *S* molecules consumption | [33, 34] |
| $S_{th1}$ | 0.8 | | Fraction of the *S* concentration that defines the threshold for the G1m/G1p transition | This paper (by data fitting) |
| $S_{th2}$ | 0.05 | | Fraction of the *S* concentration that defines the threshold for the G1p/S transition | This paper (by data fitting) |
| $CycX_{th}$ | $8x10^{-17}$ | kg | Amount of cyclin X that defines the threshold for the G2/M transition | This paper (by data fitting) |
| $C1$ | $3x10^{16}$ | $kg^{-1}$ | Phenomenological constant for the growth of mitochondria | This paper (by data fitting) |
| $C2$ | $2x10^{-18}$ | $m^{3}$ | Phenomenological constant for the increase of the cell volume | This paper (by data fitting) |
| $v_{Mit}$ | $4x10^{-22}$ | kg $s^{-1}$ | Rate of ATP consumption for mitochondria maintenance | This paper (by data fitting) |
| $ATP_{min}$ | $1.5x10^{-15}$ | kg | Minimum ATP amount for cell survival | This paper (by data fitting) |
| *S_T* | 55000 | s | Maximum duration of the S phase | [6] |
| *G2_T* | 25200 | s | Maximum duration of the G2 phase | [6] |
| $M_T_{M}$ | 1800 | s | Mean duration of the M phase | [6] |
| *PHASE_SPREAD* | 0.5 | | Maximum spread for the fluctuations in the M phase duration | This paper (by data fitting) |

[a] value measured for the equilibrium between ATP and cyclin-dependent kinases.

**Table 2.** Estimated parameters for the simulated standard tumor cell population compared to experimental measurements. All values have been taken during exponential growth.

| **Parameter** | **Simulated** | | | **Experimental** | **Ref.** |
|---|---|---|---|---|---|
| Growth rate [a] (hours$^{-1}$) | 0.035 | | | 0.0304 [b]<br>0.0352 [c] | This work |
| Doubling time[a] (hours) | 19.8 | | | 22.8 [b]<br>19.7 [c] | This work |
| | **Average** | **Min** | **Max** | **Mean±SD** | |
| Cell cycle distribution (%) | | | | | |
| G1<br>S<br>G2/M | 52.5<br>34.5<br>12.9 | 48.4<br>30.5<br>7.3 | 59.3<br>40.5<br>17.7 | 54.4 ± 2.2 [a]<br>27.5 ± 5.8 [a]<br>16.4 ± 1.7 [a] | [6] |
| ATP/Cell ($10^{-15}$ g) | 5.47 | 5.37 | 5.55 | 5.76 ± 0.73 [a]<br>4.32 ± 0.72 [b] | This work |
| Radius (μm) | 5.02 | 4.82 | 5.30 | ≈7.5 | [61] |
| Volume (μm$^3$) | 530 | 471 | 623 | 700-1500 | [61, 62] |
| Mitochondria/Cell | 220.4 | 190.6 | 266.9 | 83-677 [d] | [63] |

[a]The growth rate for both simulated and experimental cell populations was calculated by exponential fitting of growth curves. The doubling time was then calculated as log2/(growth rate)
[b]Data measured for the MOLT3 human T lymphoblastoid cell line
[c]Data measured for the Raji human B lymphoblastoid cell line
[d]Range of the number of mitochondria observed in different cell types

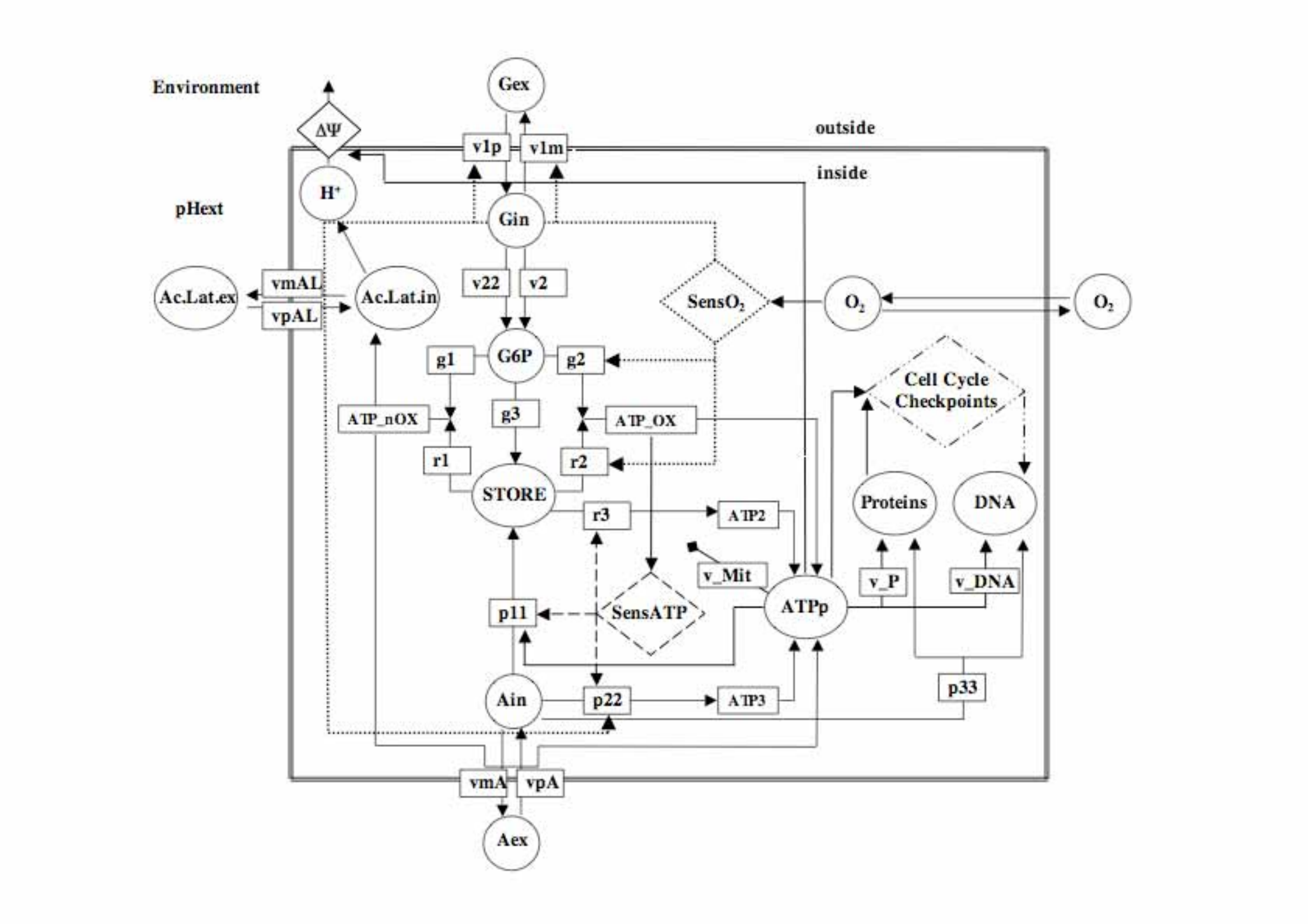

Fig.1

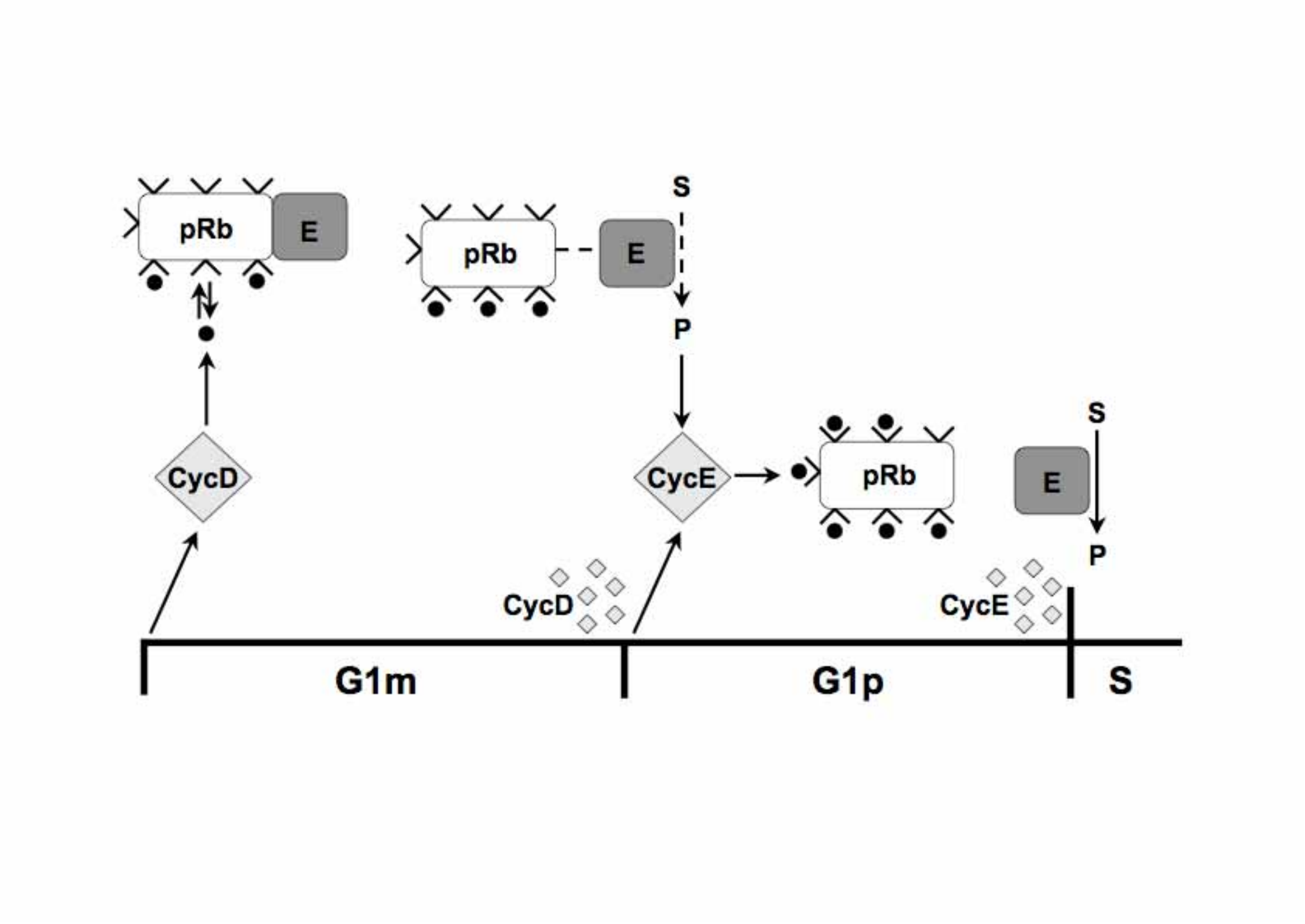

Fig.2

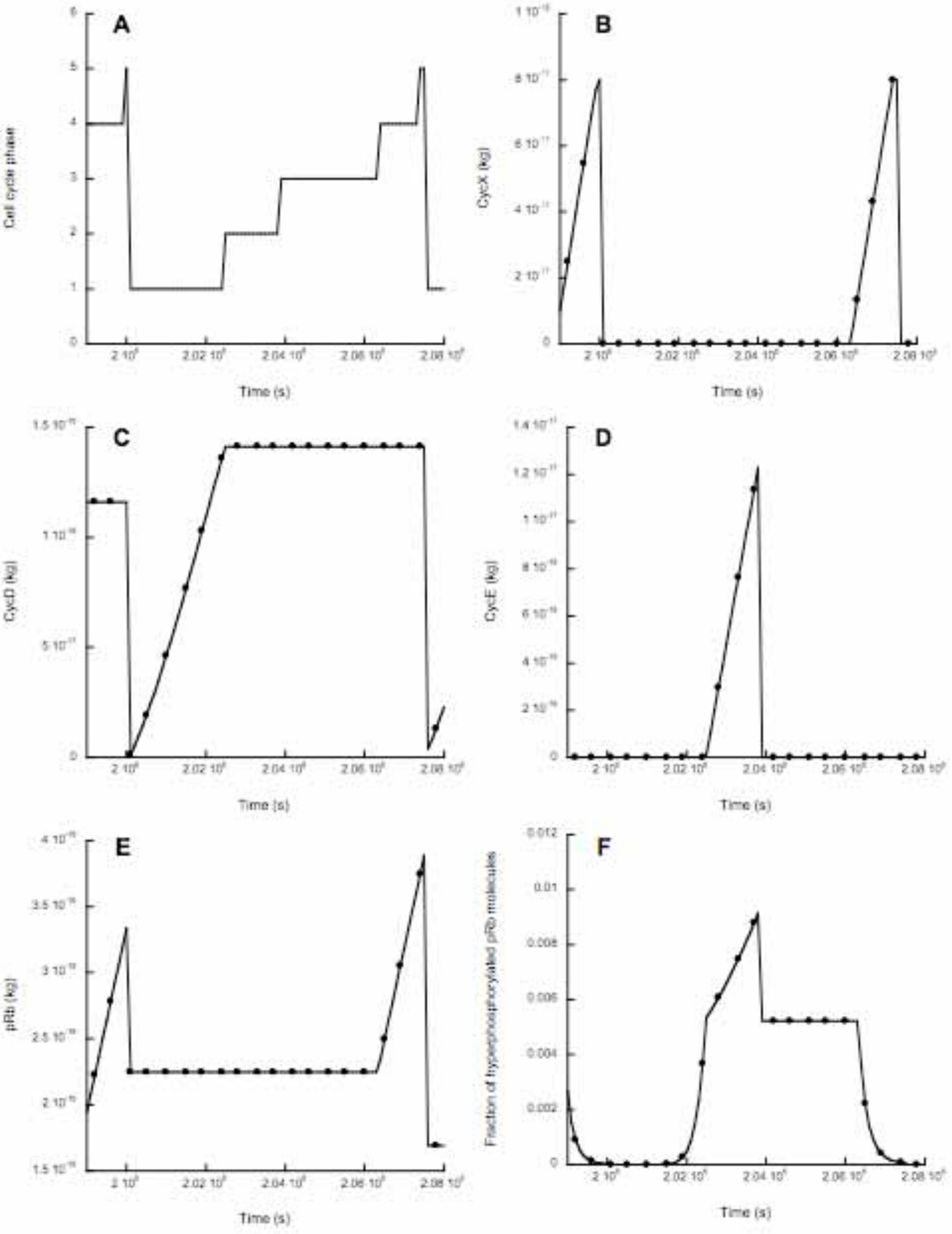

Fig.3

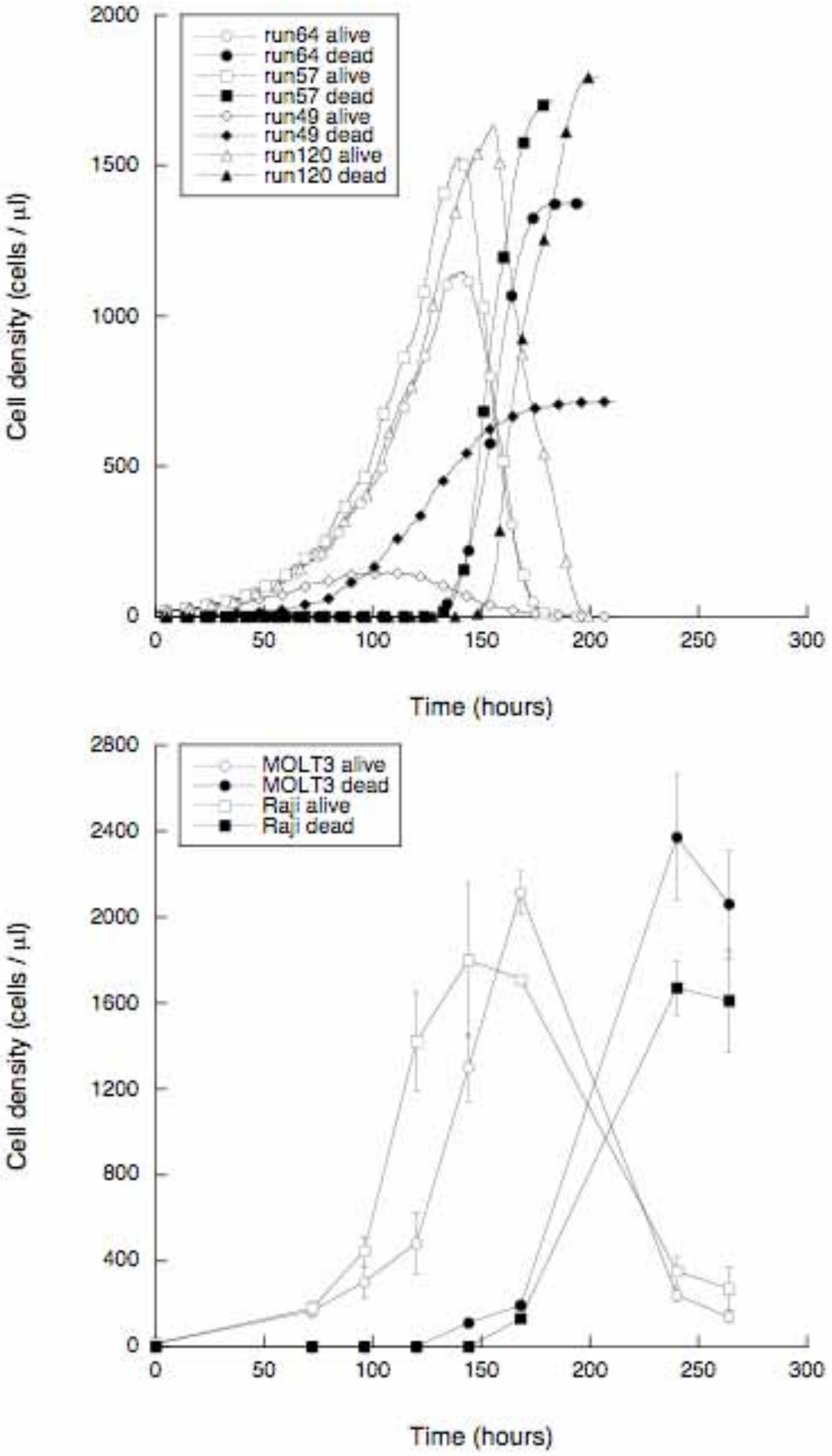

Fig.4

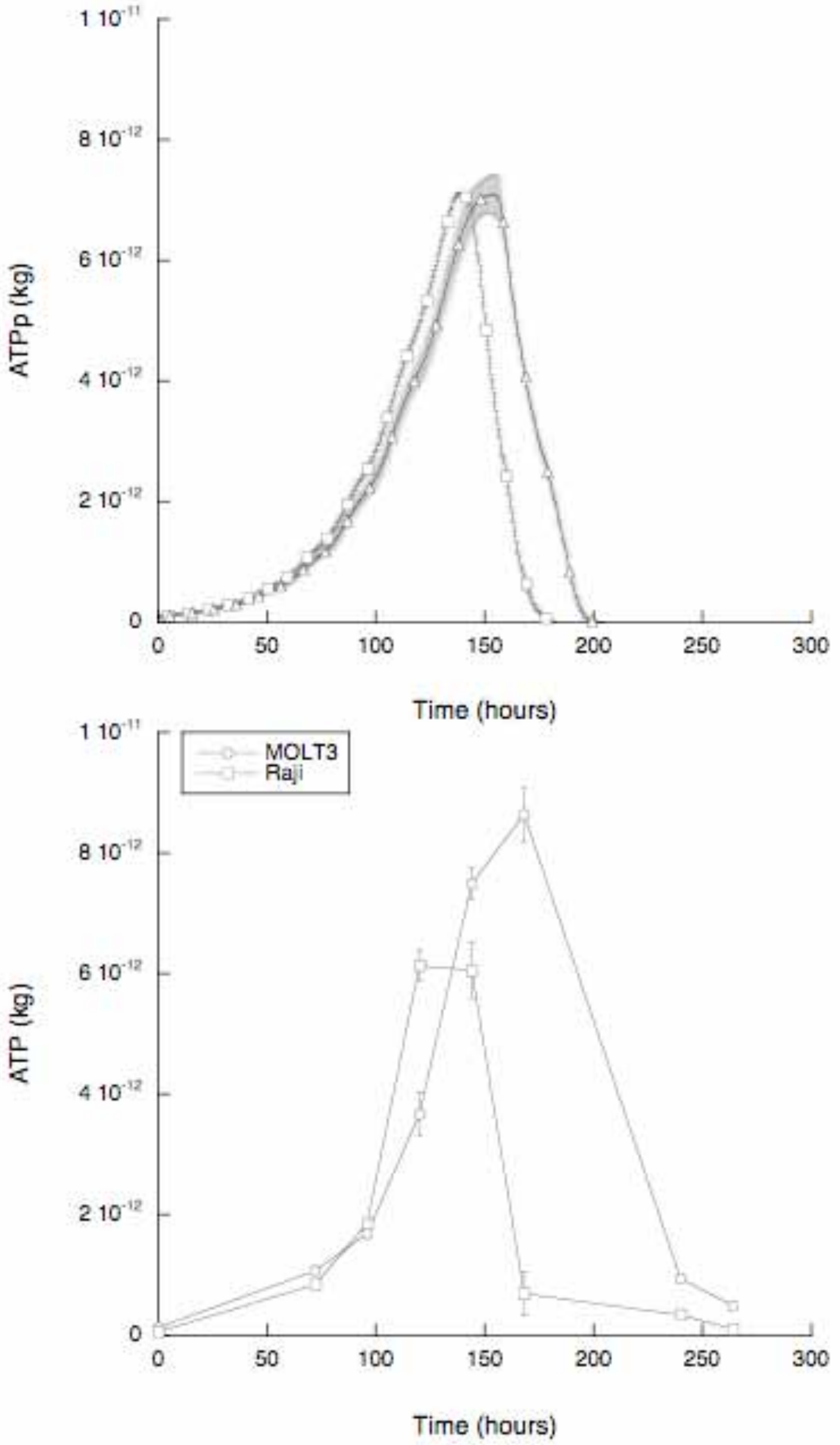

Fig.5

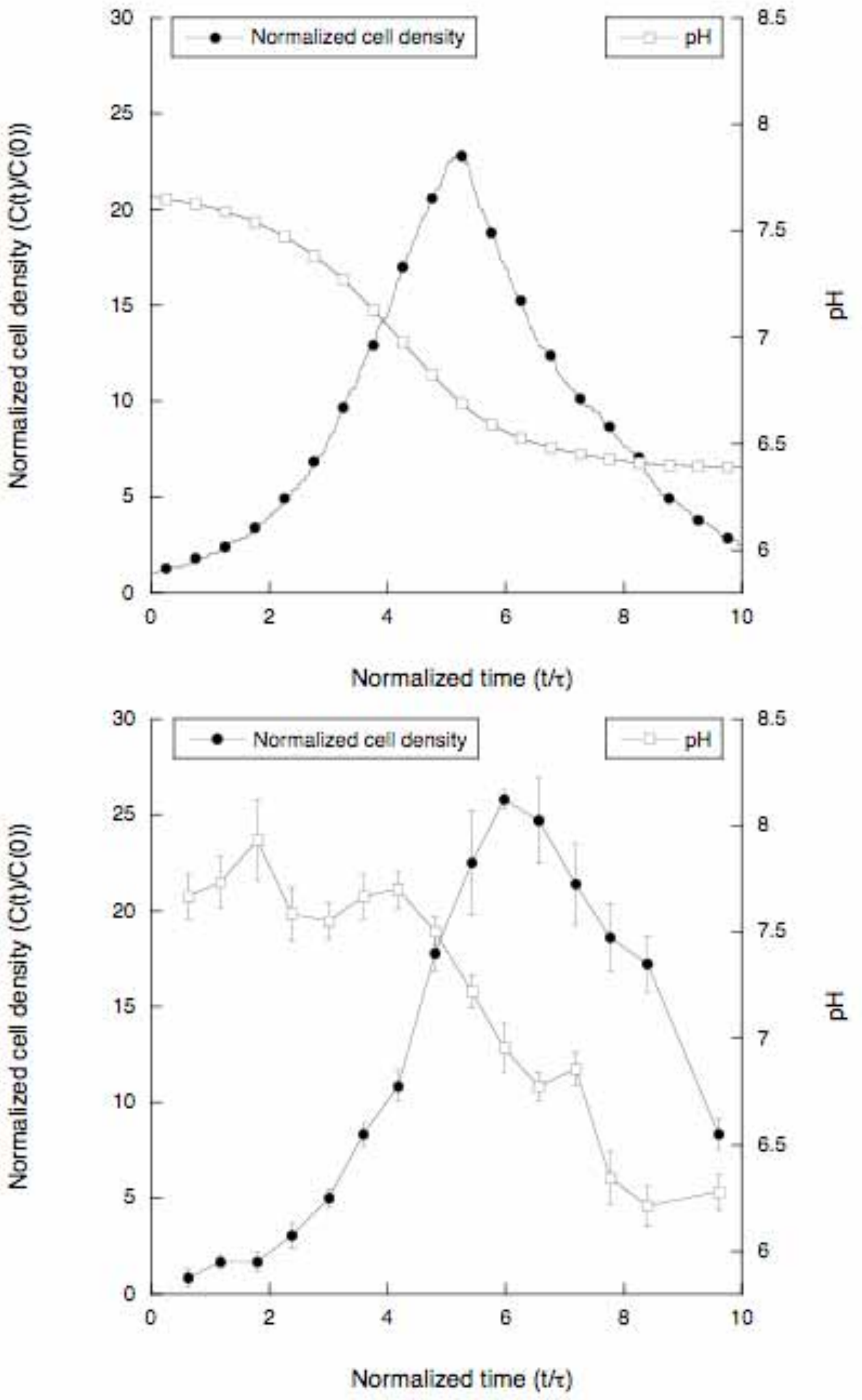
Normalized cell density
pH
Normalized cell density (C(t)/C(0))
Normalized time (t/τ)
Normalized cell density
pH
Normalized cell density (C(t)/C(0))
Normalized time (t/τ)

Fig.6

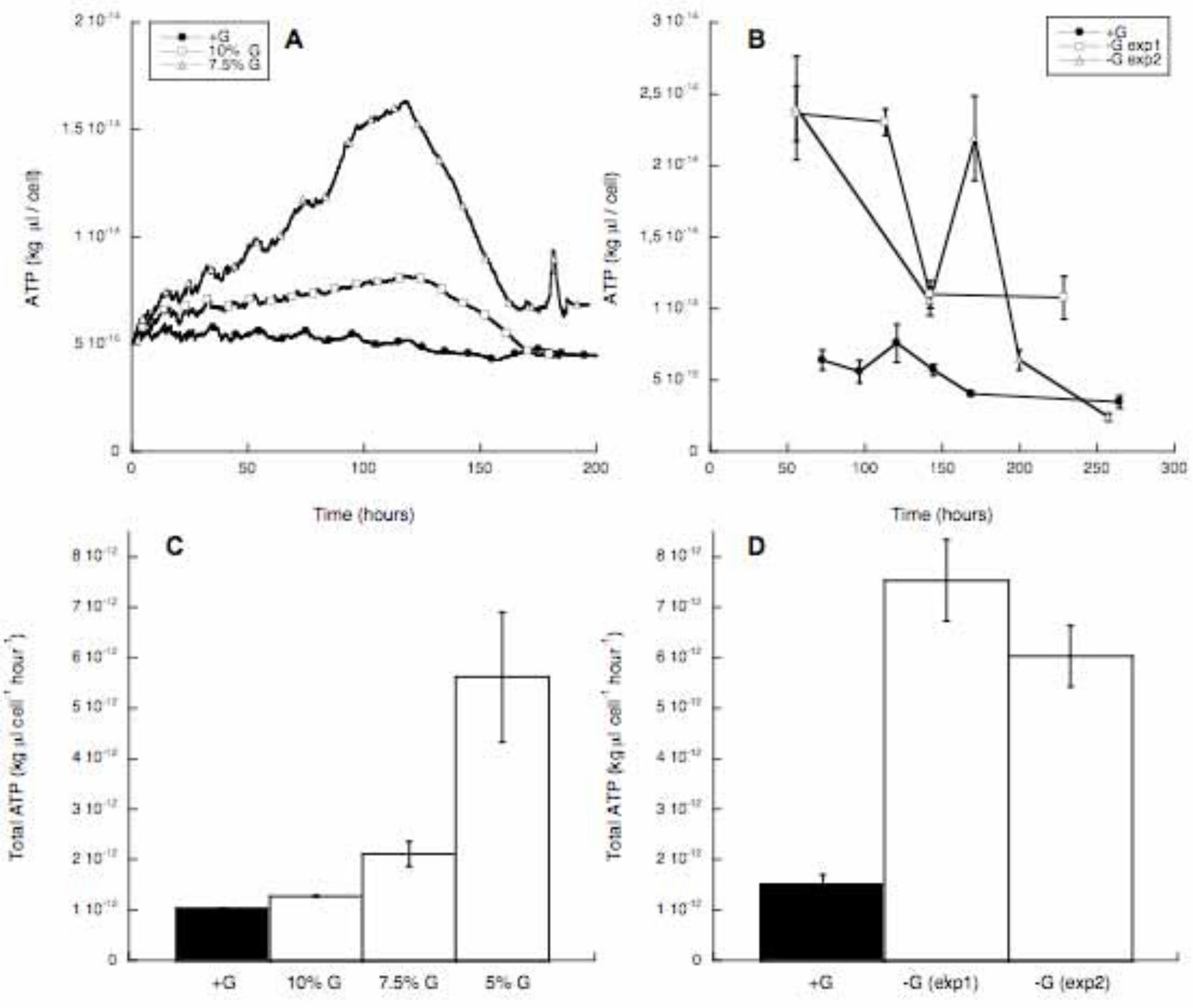

Fig.7

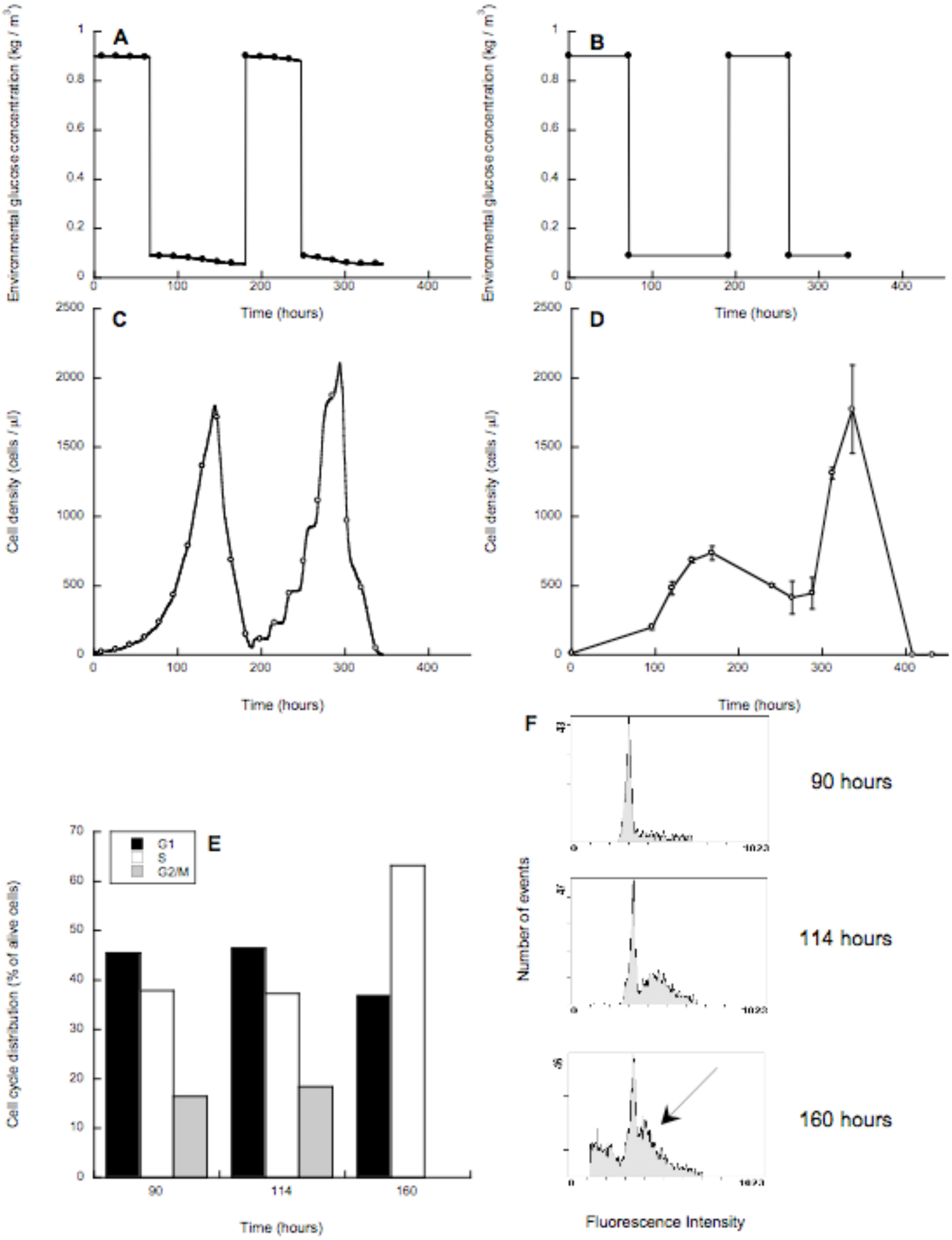

Fig.8